\documentclass[preprint2]{aastex}

\newcommand{\Rsun}{R$_\odot$}
\newcommand{\Msun}{M$_\odot$}

\newcommand{\myemail}{s.dzib@crya.unam.mx}

\slugcomment{First Draft}

\shorttitle{A VLA survey of Ophiuchus}
\shortauthors{Dzib et al.}

\usepackage{epsfig}
\usepackage{longtable}
\usepackage{amssymb}
\usepackage{lscape}

\begin{document}

\newpage

\title{The Gould's Belt Very Large Array Survey I:\\
The Ophiuchus complex}

\author{Sergio Dzib\altaffilmark{1}, 
Laurent Loinard\altaffilmark{1,2}, 
Amy J.\ Mioduszewski\altaffilmark{3}, 
Luis F.\ Rodr\'{\i}guez\altaffilmark{1,4}, 
Gisela N.\ Ortiz-Le\'on\altaffilmark{1}, 
Gerardo Pech\altaffilmark{1}, 
Juana L.\ Rivera\altaffilmark{1}, 
Rosa M.\ Torres\altaffilmark{5}, 
Andrew F.\ Boden\altaffilmark{6}, 
Lee Hartmann\altaffilmark{7}, 
Neal J.\, Evans II\altaffilmark{8}, 
Cesar Brice\~no\altaffilmark{9} and
John Tobin\altaffilmark{10}
}

\altaffiltext{1}{Centro de Radioastronom\'{\i}a y Astrof\'{\i}sica, Universidad
Nacional Aut\'onoma de M\'exico\\ Apartado Postal 3-72, 58090,
Morelia, Michoac\'an, Mexico (\myemail)}

\altaffiltext{2}{Max Planck Institut f\"ur Radioastronomie, Auf dem H\"ugel 69, 53121 Bonn, Germany}

\altaffiltext{3}{National Radio Astronomy Observatory, Domenici Science Operations Center,\\
1003 Lopezville Road, Socorro, NM 87801, USA}

\altaffiltext{4}{King Abdulaziz University, P.O. Box 80203, Jeddah 21589, Saudi Arabia}

\altaffiltext{5}{Paul Harris 9065, Las Condes, Santiago, Chile }

\altaffiltext{6}{Division of Physics, Math, and Astronomy, California Institute of Technology, 1200 E California Blvd., 
Pasadena, CA 91125, USA}

\altaffiltext{7}{Department of Astronomy, University of Michigan, 500 Church Street, Ann Arbor, MI 48105, USA}

\altaffiltext{8}{Department of Astronomy, The University of Texas at Austin, 1 University Station, C1400, Austin, 
TX 78712, USA}

\altaffiltext{9}{Centro de Investigaciones de Astronom\'ia, M\'erida, 5101-A, Venezuela}

\altaffiltext{10}{National Radio Astronomy Observatory, Charlottesville, VA 22903}

\begin{abstract}
We present large-scale ($\sim$ 2000 square arcminutes), deep ($\sim$ 20 $\mu$Jy), 
high-resolution ($\sim$ 1$''$) radio observations of the Ophiuchus star-forming complex
obtained with the Karl G.\ Jansky Very Large Array at $\lambda$ = 4 and 6 cm. In total,
189 sources were detected, 56 of them associated with known young stellar sources, 
and 4 with known extragalactic objects; the other 129 remain unclassified, but most 
of them are most probably background quasars. The vast majority of the young stars detected 
at radio wavelengths have spectral types K or M, although we also detect  4 objects 
of A/F/B types and 2 brown dwarf candidates. At least half of these young stars are 
non-thermal (gyrosynchrotron) sources, with active coronas characterized by high 
levels of variability, negative spectral indices, and (in some cases) significant circular 
polarization. As expected, there is a clear tendency for the fraction of non-thermal 
sources to increase from the younger (Class~0/I or flat spectrum) to the more evolved 
(Class~III or weak line T Tauri) stars. The young stars detected both in X-rays and at 
radio wavelengths broadly follow a G\"udel-Benz relation, but with a different normalization 
than the most radio-active types of stars. Finally, we detect a $\sim$ 70 mJy compact 
extragalactic source near the center of the Ophiuchus core, which should be used as 
gain calibrator for any future radio observations of this region.
\end{abstract}

\keywords{astrometry ---magnetic fields --- radiation mechanisms: non--thermal --- 
radio continuum: stars --- techniques: interferometric}

\section{Introduction}
Gould's Belt (see Poppel 1997 for a recent review) is a flattened local Galactic structure, 
about 1 kpc in size, that contains much of the recent star-forming activity in the Solar
neighborhood. In particular, it comprises the nearest sites of active star-formation -- i.e.\ 
the Taurus, Ophiuchus, and Orion molecular complexes. This paper is the first in a series 
that will describe the results of deep, large-scale radio observations of these nearby 
regions of star-formation; it will focus on the Ophiuchus complex.

A combination of several properties conspire to make Ophiuchus one of the most 
interesting targets for star-formation studies (see Wilking et al.\ 2008 for a recent
review). Its proximity (120 pc; Loinard et al.\ 2008) ensures high linear spatial 
resolution, and enables the detection of fainter objects (e.g.\ brown dwarfs) than 
in other regions. In addition, the stellar population associated with the central 
dark cloud Lynds 1688 (often called the Ophiuchus core; see Figure 1) provides 
us with an example of a young ($\tau$ $<$ 0.5 Myr; Wilking et al.\ 2008) stellar 
cluster that probes a mode of star-formation intermediate between the more 
isolated situation exemplified by Taurus, and the more clustered mode typically 
found in higher mass regions. As a consequence, Ophiuchus has been one of 
the best-studied regions of star-formation. In the last decade alone, very detailed
surveys of its stellar population have been obtained in X-rays (Imanishi et al.\ 
2001; Gagn\'e et al.\ 2004; Ozawa et al.\ 2005; Sciortino et al.\ 2006), near-infrared 
(Alves de Oliveira et al.\ 2010; Barsony et al.\ 2012), and mid-infrared (Padgett et 
al.\ 2008). The properties of the circumstellar disks around these young stars have 
been studied in the mid-infrared (Kessler-Silacci et al.\ 2006; Geers et al.\ 2006; 
Lahuis et al.\ 2007) and millimeter/sub-millimeter regimes (Andrews \& Williams 
2007). Finally, the distribution of the interstellar material has also been studied 
in detail thanks to millimeter spectroscopy (Ridge et al.\ 2006), sub-millimeter 
continuum observations (Motte et al.\ 1998; Johnstone et al.\ 2000; Stanke et 
al.\ 2006; Young et al.\ 2006), and far-infrared imaging (Padgett et al.\ 2008). 

Numerous radio observations have also been obtained, but the characterization
of the Ophiuchus complex at radio wavelengths remains significantly less systematic
than in other frequency bands. Following the early work by Brown \& Zuckerman (1975) 
and Falgarone \& Gilmore (1981), the first large-scale observations were obtained with 
the Very Large Array (VLA) in its C and D configurations at 1.4 and 5 GHz by Andr\'e 
et al.\ (1987) and Stine et al.\ (1988). While most of the detected radio sources were 
background extragalactic objects, roughly a dozen could be confidently identified with 
young stars in the Ophiuchus cluster. The next major step came with the deep 5 GHz
VLA observations of the Ophiuchus core, again in the C and D configuration, reported 
by Leous et al.\ (1991). These observations targeted two specific fields, but reached 
a noise level several times better than that of previous observations. A total of 35 
compact radio sources were identified, most of them associated with young stellar
objects (Wilking et al.\ 2001). More recently, Gagn\'e et al.\ (2004) reported the detection at 
6 cm of a dozen additional compact sources, but it is not clear what fraction of these 
sources might be associated with young stars. Finally, a number of observations 
(with better sensitivity and/or higher angular resolution) have been obtained toward 
specific objects (particularly the prototypical Class~0/I sources IRAS~16293--2422,
VLA~1623, and YLW15 --Wootten 1989; Chandler et al.\ 2005; Loinard et al.\ 2007, 
2013;  Andr\'e et al.\ 1993; Ward-Thompson et al.\ 2011; Girart et al.\ 2000, 2004).

The analysis of the radio properties of young stars in Ophiuchus has helped clarify the 
origin of the radio emission produced by young stellar objects in general. For sources 
with flat, Class I or Class II infrared spectral energy distributions, the radio emission (at
$\nu$ $\lesssim$ 30 GHz) tends to be of thermal bremsstrahlung (free-free) origin,\footnote{The 
Class~0 source IRAS~16293--2422 is somewhat of an oddity where thermal dust emission appears 
to remain the dominant emission process well into the centimeter regime (Loinard et al.\ 2013 
and reference therein). This, however, is a very unusual situation, presumably partly related 
to the large amount of dense gas and dust around IRAS~16293--2422; we do not expect to find any 
source of this type in the present survey.} and to trace the dense central base of ionized winds 
(e.g.\ Rodr\'{\i}guez 1999). This type of emission is characterized by a spectral index $\alpha$ 
(defined such that the flux depends on frequency as S$_{\nu}$ $\propto$ $\nu^{\alpha}$) 
between --0.1 and $+$2 ($\alpha$ = $+$0.6 for the classical case of a partially optically thick 
isotropic wind; Panagia 1973), and only presents moderate variability ($\lesssim$ 20\%). In 
contrast, for Class III sources, the emission is generally non-thermal (gyrosynchrotron) and 
related to magnetic activity near the stellar surface. This kind of emission often (but not always) 
exhibits strong variability and some level of circular polarization (Dulk 1985). The spectral 
index for gyrosynchrotron emission depends on the energy distribution of the electrons, and 
can vary between $-$2 and $+$2 (Dulk 1985), although a slightly negative index occurs in many cases. 
Furthermore, non-thermal radiation mechanisms produce high brightness emission (T$_b$ $\gtrsim$ 
10$^{6-7}$ K) confined to a very compact region ($\lesssim$ 10 \Rsun). As a consequence, it is 
detectable even in very long baseline interferometry (VLBI) radio experiments, as shown by 
Lestrade et al.\ (1991), Phillips et al.\ (1991) or Andr\'e et al.\ (1992). It should be mentioned that 
the separation between thermal (free-free) and non-thermal (gyrosynchrotron) emission at the Class II/III 
boundary is not a sharp one as a number of Class II and even a few Class I sources have been 
found to emit non-thermal radiation (e.g.\ Forbrich et al.\ 2007; Dzib et al.\ 2010; Deller et al. 2013). 

In the present article, we report on new radio observations of the Ophiuchus complex
which largely surpass previously published studies thanks to a combination of high
sensitivity, sub-arcsecond angular resolution, and large field of view. These data will 
be used both to discuss the population of radio sources in Ophiuchus, and the relation
between radio properties of young stellar objects and their characteristics at other
wavelengths. The observations are described in Section 2; the results are 
presented in Section 3, and analyzed in Sections 4 and 5. Our conclusions are provided 
in Section 6.

\section{Observations}

The observations were obtained with the Karl G. Jansky Very Large Array (VLA)
of the National Radio Astronomy Observatory (NRAO) in its CnB and B configurations. 
Two frequency sub-bands, each 1 GHz wide, and centered at 4.5 and 7.5 GHz, 
respectively, were recorded simultaneously. The observations were obtained on three 
different epochs (February 17/19; April 3/4, and May 4/6, 2011) typically separated from 
one another by a month. This dual frequency, multi-epoch strategy was chosen to enable 
the characterization of the spectral index and variability of the detected sources, and to
help in the identification of the emission mechanisms (thermal vs. non-thermal). 

\begin{figure*}[!ht]
\begin{center}
 \includegraphics[width=0.95\textwidth,angle=0]{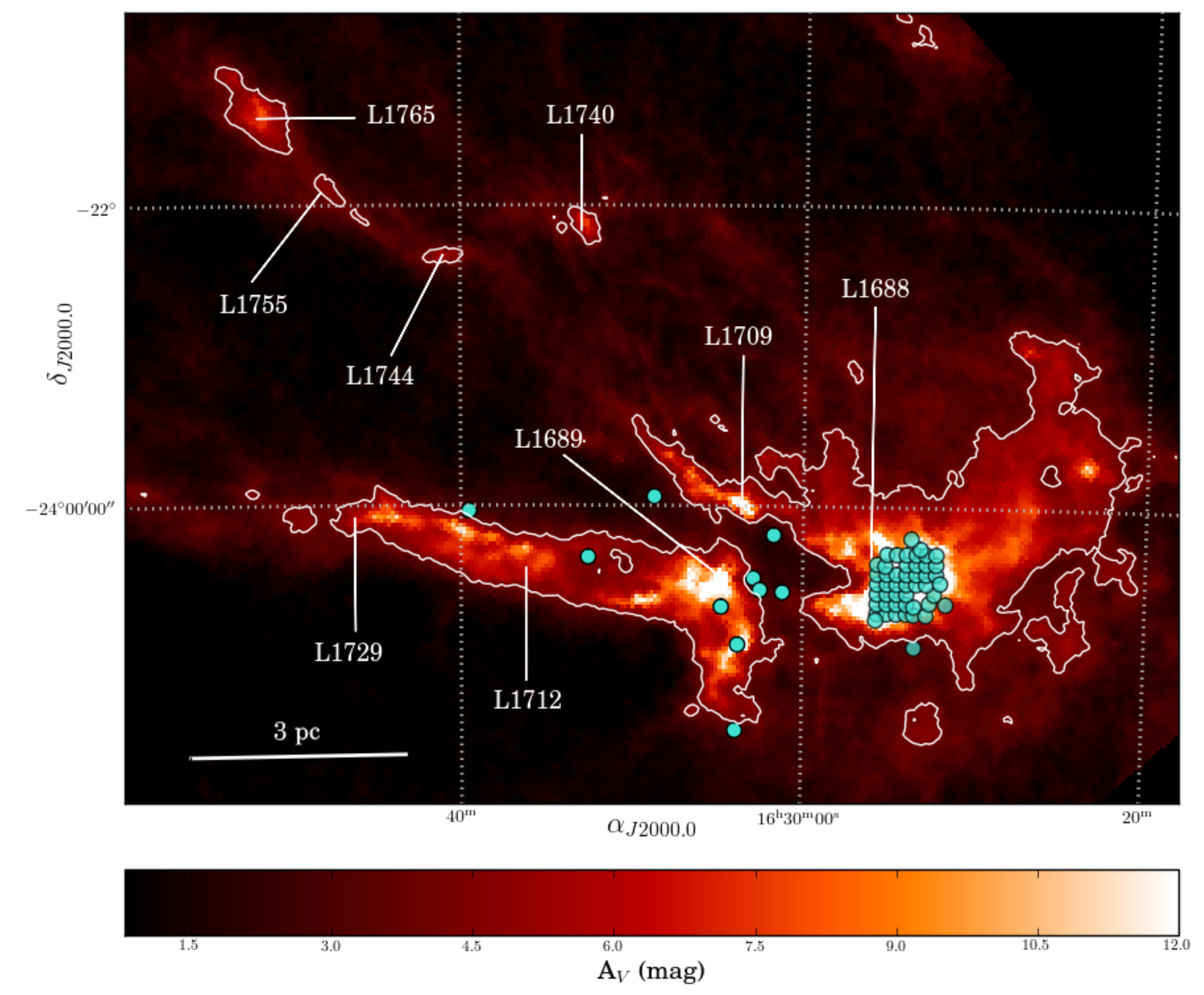}
\end{center}
 \caption{Extinction map of the Ophiuchus complex obtained as part of the COMPLETE project (Ridge et al.\ 2006)
 based on the 2MASS data (Skrutskie et al.\ 2006). The Lynds dark clouds in the regions are indicated, and a linear
 distance is provided (assuming a distance of 120 pc to the entire region --Loinard et al.\ 2008). The turquoise circles
 indicate the areas mapped with the VLA for the survey presented here. The diameter of each circle is 6$'$ and 
 corresponds to the primary beam of the VLA at 7.5 GHz. Note that the field of view, and therefore also the total mapped
 area, at 4.5 GHz is significantly larger.}
\label{fig:map+fields}
\end{figure*}

Since our aim was to examine the distribution and properties of radio sources in the 
Ophiuchus complex in a statistically meaningful fashion, it was important to 
systematically map a large area of the complex, including those regions known 
to harbor a high density of young stars. The region of highest stellar density is 
associated with the dark cloud Lynds 1688 (the Ophiuchus core). We mapped this 
area, which contains over 400 young stellar objects, using a mosaic of 47 VLA 
pointings (see Figure \ref{fig:map+fields}). Ten additional pointings were selected to 
cover regions associated with other dust clouds (particularly L1689, but also L1709, 
L1712, and L1729; see Figure \ref{fig:map+fields}) located to the east of the Ophiuchus 
core (these clouds are known collectively as the Ophiuchus {\it streamers}).

The FWHM of the primary beam (i.e.\ the field of view) of the VLA has a diameter of 10$'$ at 4.5 
GHz, and 6$'$ at 7.5 GHz. As a consequence, the ten individual fields targeting the 
streamers cover an area of 785 square arcminutes at 4.5 GHz and 283 square
arcminutes at 7.5 GHz. The spacing between the individual pointings of the mosaic 
observed toward L1688 follows a somewhat irregular pattern chosen to optimize 
the compromise between uniform sensitivity and inclusion of the largest possible
number of known young stars. The total area covered by these fields is 1185 square 
arcminutes at 4.5 GHz and 878 square arcminutes at 7.5 GHz. 

Each observing session was organized as follows. The standard flux calibrator 
3C~286 was first observed for $\sim$10 minutes. We subsequently spent one 
minute on the phase calibrator J1626-2951 followed by a series of three target
pointings, spending three minutes on each. This phase calibrator/target sequence
was repeated until all target fields were observed. Thus, three minutes were
spent on each target field for each epoch. The data were edited, calibrated, and 
imaged in a standard fashion using the Common Astronomy Software Applications 
package (CASA).

In the 10 individual fields associated with the streamers, the noise level reached 
for each epoch was 61 and 52 $\mu$Jy beam$^{-1}$ at 4.5 GHz and 7.5 GHz, 
respectively. For the mosaicked region, on the other hand, a nearly uniform noise
level of 26 $\mu$Jy beam$^{-1}$ is obtained at both frequencies. The frequency
independence of the noise is the result of two effects that compensate each 
other: While the noise in individual fields is somewhat better at 7.5 GHz, the larger 
field of view at lower frequency results in a larger overlap between fields which 
reduces the noise in the final mosaic. To produce images with improved sensitivity,
the three epochs were combined, resulting in noise levels of 30 and 25 $\mu$Jy 
beam$^{-1}$ at 4.5 GHz and 7.5 GHz, respectively, for the individual (streamers)
fields, and 17 $\mu$Jy beam$^{-1}$ at both frequencies for the mosaic. The 
angular resolution of the observations is of order 1$''$. 

To test for circular polarization we produced images of the $V$ Stokes parameter 
in the inner quarter (in area) of the primary beam at each frequency. At larger distances
from the field center, polarization measurements are unreliable as beam squint 
(the separation of the $R$ and $L$ beams on the sky) can create artificial circular
polarization signals.

\section{Results}

The first step is, of course, to identify sources in our observations. This was
done using the images corresponding to the concatenation of the three epochs,
which provide the highest sensitivity. The criteria used to consider a detection 
as firm were: (i) sources with reported counterparts and a flux larger than four 
times the $\sigma$ noise of the area, or (ii) sources with a flux larger than five 
times the $\sigma$ noise of the area and without reported counterparts. From
this, a total of 189 sources were detected (see Table \ref{tab:sources}). To reflect 
the fact that these sources were found as part of the {\it Gould's Belt 
Very Large Array Survey}, a source with coordinates {\it hhmmss.ss$-$ddmmss.s} 
will be named GBS-VLA J{\it hhmmss.ss$-$ddmmss.s}.

The flux of each source at 4.5 and 7.5 GHz are given in columns 2 and 4 of Table 
\ref{tab:sources}). Two sources of uncertainties on the fluxes are included: (i) 
the error that results from the statistical noise in the images, and (ii) a systematic
uncertainty of 5\% resulting from possible errors in the absolute flux calibration.
An estimation of the radio spectral index of each source (given in column 6 of 
Table \ref{tab:sources}) was obtained from the fluxes measured in each 
sub-band (at 4.5 and 7.5 GHz). To calculate the errors on the spectral indices, 
the two sources of errors (statistical and systematic) on the flux at each frequency 
were added in quadrature and the final error was obtained using standard error
propagation theory.

Once the sources were identified in the concatenated images, we searched
for them in the images obtained from the individual epochs.\footnote{We also
searched the individual epochs for objects that might be present there although
they are not detected in the averaged data, but found no such object.} An estimate 
of the level of variability of the sources was measured by comparing the fluxes 
measured at the three epochs. Specifically, we calculated, for each source and at 
each frequency, the difference between the highest and lowest measured fluxes, 
and normalized by the maximum flux. The resulting values, expressed in 
percent, are given in columns 3 and 5 of Table \ref{tab:sources}. Circular polarization
was confidently detected in 7 of our targets (Table \ref{tab:polz}).

Having identified the radio sources in the region mapped, our next step was to try 
to determine which type of object they are associated with. In our specific case, 
the two overwhelmingly dominant possibilities are young stars and extragalactic 
sources.\footnote{Although we cannot fully rule out the possibility that other
objects might contaminate the sample (a planetary nebula, for instance), the 
probability of such an occurrence is very low.} We searched the literature for
previous radio detections, and for counterparts at X-ray, optical, near- and 
mid-infrared wavelengths. The search was done in SIMBAD, and accessed
all the major catalogs (listed explicitly in the footnote of Tab.\ \ref{tab:counterparts}). 
Note that the Spitzer c2d catalog includes cross-references to other major
catalogs which were taken into account in our counterpart search. We considered
a radio source associated with a counterpart at another wavelengths if the separation
between the two was below the combined uncertainties of the two datasets. This was
about 1.5 arcsec for the optical and infrared catalogs, but could be significantly larger
for some of the radio catalogs (for instance, the NVSS has a positional uncertainty of about 5 arcsec).
We found that only 76 of the sources detected here had previously been reported at radio 
wavelengths (column 7 of Table 
\ref{tab:counterparts}), while the other 113 are new radio detections. On the other 
hand, we found a total of 100 counterparts at other wavelengths. Note that there is
a significant number of sources that were previously known at radio wavelengths 
{\bf and} have known counterparts at other frequencies. As a consequence, the 
number of sources that were previously known (at any frequency) is 134, while 55 
of the sources in our sample are reported here for the first time. 

Based on their optical/infrared characteristics, 2 of the 100 sources with counterparts are 
classified in the literature as extragalactic sources, while 55 are classified as young stellar 
objects (YSO; see column 8 of Table  \ref{tab:counterparts}). One additional object (GBS-VLA 
J162626.31$-$242430.3, also known as VLA 1623B) is associated with the 
well-known Class~0 source VLA 1623, but it is still debated whether it is an 
outflow knot feature, or a protostellar object (see the discussion by Ward-Thompson
et al.\ 2011; and section 5.1 below). The remaining 42 radio sources with known counterparts at other
wavelengths are, to our knowledge, not classified in the literature. On the other hand, two sources 
(GBS-VLA J162615.67-243421.2 and GBS-VLA J162626.03-244923.7) have been classified
in the literature as extragalactic, on the basis of their radio properties alone. 

\begin{figure*}[b!]
\begin{center}
\includegraphics[width=0.90\textwidth,angle=0]{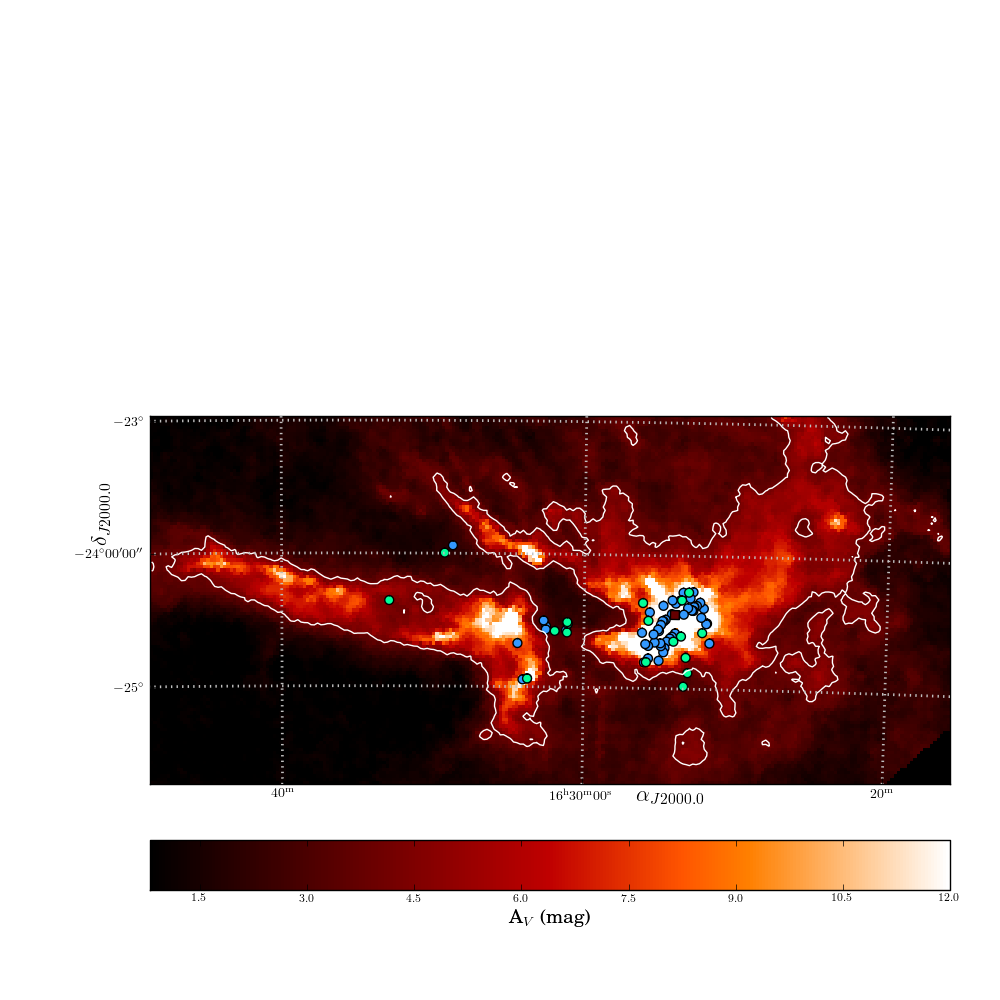}
\end{center}
\caption{Location of the YSO (blue circles) and candidate YSO (green circle) onto the
extinction map of the Ophiuchus complex. The position of the new calibrator detected here
(GBS-VLA J162700.00-242640.3) is indicated as a brown square in the Ophiuchus core.}
\label{fig:smap}
\end{figure*}

To summarize, a total 134 of our 189 sources were previously known (76 at radio 
wavelengths, and 100 at other frequencies, with some overlap between the two 
sub-samples). Of these 134 sources, 56 are classified as YSO and 4 as extragalactic;
the other 74 are not classified. Given the existing deep near and mid-infrared 
observations of Ophiuchus, it is unlikely that a major population exists of 
unidentified young stars. In consequence, we argue that most of these unclassified 
sources are extragalactic. For the same reason, most of the 55 objects detected here 
for the first time are likely background sources. We note, however, that 18 of the
129 unclassified objects (55 identified here for the first time and 74 previously known
at radio wavelengths) are compact, have a positive spectral index, or exhibit high
variability (Table \ref{tab:possibleYSOs}). Since these latter two properties are not 
expected of quasars (which are certainly variable, but usually not strongly on such 
short timescale --e.g.\ Hovatta et al.\ 2008), but would be natural characteristics of 
young stars, we argue that a small population of YSO might be present among the 
unclassified sources. This population could account for, at most, 15\% of the unclassified 
sources, and possibly significantly less. The distribution of both the YSO and candidate 
YSO in our radio sources are shown in Figure \ref{fig:smap}.

It is interesting that 35 of the sources reported here are only detected at radio frequencies and
with Spitzer (with no detection at any other wavelengths). These sources are usually not
classified, and might be either young stellar sources, or foreground/background objects.
It will be useful to study this population further.

\section{Discussion}

\subsection{Background Sources}

It is clear from Section 3 that a large fraction of the radio sources detected here 
are background objects. To examine their statistics, we will concentrate on the core
region (for which we have a continuous coverage at uniform sensitivity) and on the
observations at 4.5 GHz, which are more appropriate than those at 7.5 GHz for extragalactic
objects, since those usually have negative spectral indices. Fomalont et al.\ (1991) showed 
that the source counts at 5 GHz are appropriately described by

\begin{displaymath}
 \left(\frac{N}{arcmin^2}\right)=0.42\pm0.05\left(\frac{S}{30 \mu Jy}\right)^{-1.18\pm0.19}.
\end{displaymath}

\noindent
According to those counts, the number of sources brighter than 150 $\mu$Jy (the minimum flux 
of the sources that we detected; see Table \ref{tab:sources}) expected in the 1185 square 
arcminutes covered by our Ophiuchus core observations is 75 $\pm$ 9. The number of
objects classified as extragalactic in the core is 3, and there are 79 unclassified sources in that 
region. As we mentioned earlier, most of these unclassified source are likely extragalactic, with
only a small contribution (of at most 15\%) of YSO. This means that about 67 of the unclassified 
sources in the core are extragalactic, and that the total number of observed extragalactic sources 
in that region is about 70, in excellent agreement with the count predicted by Fomalont et al.\ (1991). 
Note that the counts would in fact also be consistent with the possibility that {\bf all} the unidentified 
sources are extragalactic. 

\subsection{Radio properties of the YSO population}

\begin{figure*}[b!]
\begin{center}
\includegraphics[width=0.60\textwidth,angle=0]{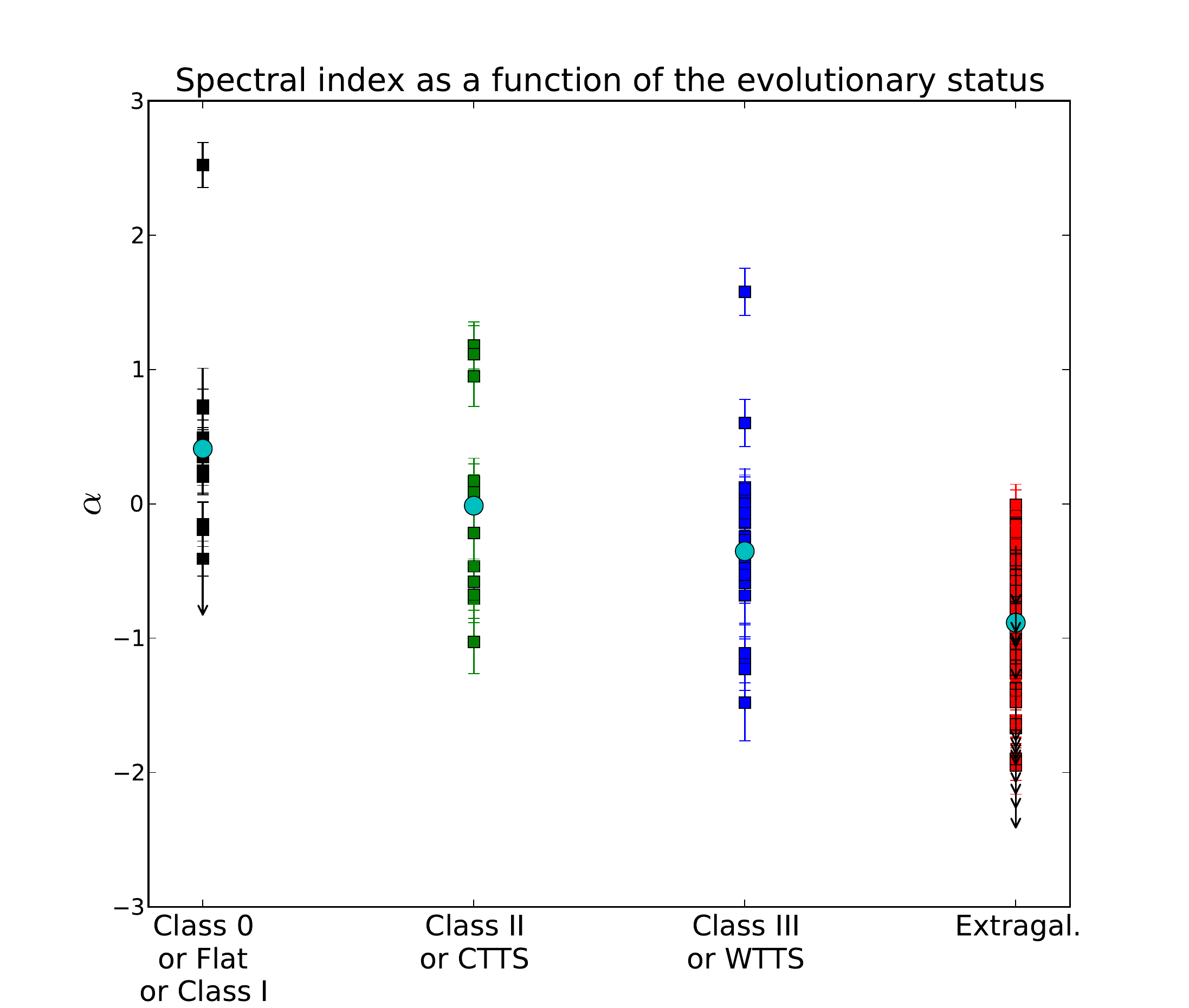}
\end{center}
\caption{Spectral index as a function of the YSO evolutionary status.
Extragalactic objects are also displayed for reference. The individual sources
are shown with their error bars, and the turquoise circles indicate the mean spectral
index for each category.}
\label{fig:AC}
\end{figure*}

In section 3, we mentioned that 56 of the radio sources detected here are associated with young 
stars. The spectral type and evolutionary status for most of these objects are known (see Table 
\ref{tab:yso}) and can be compared to their radio properties. In Figure \ref{fig:AC}, the radio spectral 
index is plotted as a function of evolutionary status. There is a clear tendency for more evolved
YSOs to have a smaller (i.e.\ more negative) spectral index. The younger (Class 0, flat spectrum,
and Class I) sources have a mean spectral index of order 0.5, indicating that the dominant 
emission process is somewhat optically thick free-free emission. The older (Class II and III) sources, 
on the other hand, have a slightly negative mean spectral index, suggestive either of optically thin 
free-free emission or of gyrosynchrotron radiation (see Section 1). In particular, there is a significant 
population of Class II/III objects with $\alpha$ $<$ --0.1, and which are most likely non-thermal emitters. 
Note, however, that the boundaries are not sharp since some very young objects have slightly 
negative indices, while one of the more evolved YSOs reach a spectral index above one. As expected, 
the extragalactic objects in our sample typically have negative spectral indices, with a mean value of
order --1. 

\begin{figure*}[b!]
\begin{center}
\includegraphics[width=0.80\textwidth,angle=0]{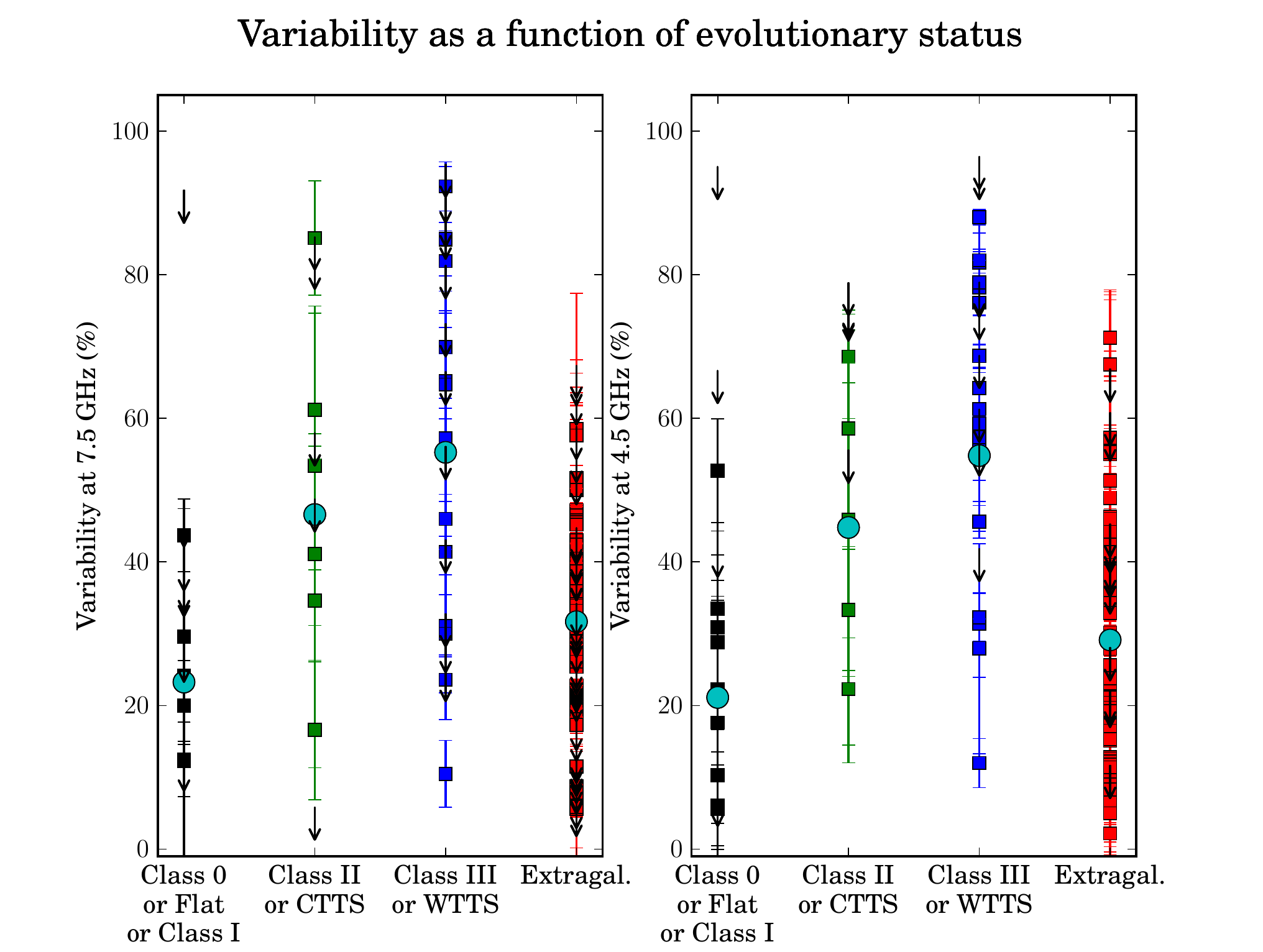}
\end{center}
\caption{Variability at 7.5 GHz (left) and 4.5 GHz (right) as a function of YSO 
evolutionary status. Extragalactic objects are also displayed for reference. The 
individual sources are shown with their error bars, and the turquoise circles indicate 
the mean variability for each category.}
\label{fig:VCl}
\end{figure*}

\begin{figure*}[b!]
\begin{center}
\includegraphics[width=0.80\textwidth,angle=0]{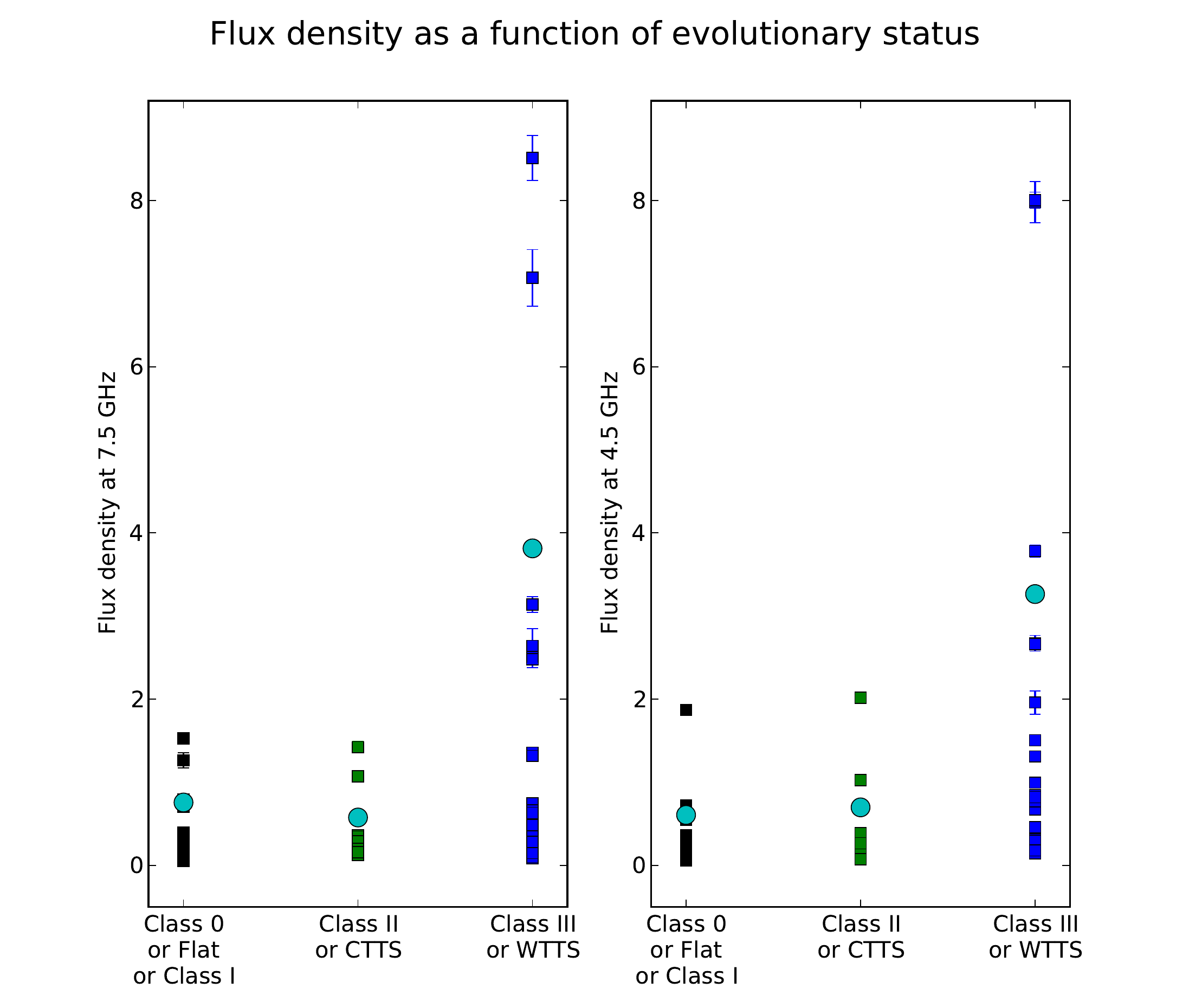}
\end{center}
\caption{Radio flux at 7.5 GHz (left) and 4.5 GHz (right) as a function of YSO 
evolutionary status. As in the previous figures, the individual sources are shown 
with their error bars, and the black circles indicate the mean  flux for each category.}
\label{fig:fluxvsevol}
\end{figure*}

The radio variability is shown as a function of evolutionary status in Figure \ref{fig:VCl}. It is clear
that younger sources are, on average, less variable than their more evolved counterparts. Since 
non-thermal emitters are often strongly variable, the measured increase in variability confirms the 
conclusion drawn above from the spectral index analysis that there is a significant population of 
non-thermal sources among the Class II/III sources in our sample. A final piece of information 
supports that conclusion: The radio flux appears to also be, on average, higher for
more evolved sources (Figure \ref{fig:fluxvsevol}), particularly from objects of Class III as compared
to those of Class II. Since younger sources experience more intense outflow activity, they will be 
stronger thermal emitters than more evolved objects (this is consistent with the observed variability 
and spectral index trends described above). The stronger average radio flux for more evolved sources 
must, therefore, indicate that a different mechanism dominates the radio emission as young stars 
age. The most natural candidate is, again, gyrosynchrotron emission. It is important to point out, 
however, that young YSOs might well intrinsically emit as much non-thermal radiation as their older 
siblings, but because of their dense ionized winds, such emission might be absorbed by the optically 
thick free-free emission (e.g. Forbrich et al.\  2007; Deller et al.\ 2013; see also Section 5.3). 

The previous discussion shows that there exists a significant population of non-thermal radio
sources in our sample of detected YSOs. From Table \ref{tab:yso}, we can see that 25 of the 55
young stars in our sample (45\%) either are polarized, or have high variability and a negative
spectral index. These source are almost certainly non-thermal and, as expected, they are mostly 
somewhat evolved --64\% of them are Class III and/or WTTS. There might be an even larger 
population of non-thermal YSOs in our sample since those sources which are either highly variable 
but with positive spectral index, or steady but with a  negative spectral index might also fall in
that category. Finally, if any of the unclassified sources given in Table \ref{tab:possibleYSOs} are 
indeed young stellar sources, they must also be non-thermal. 

A final trend must be mentioned here. Out of the 36 detected YSO with known
spectral type, 31 fall in the range K0 to M5 (Figure 6), while no G type star, 
and only one F, two A, and one B stars are detected. This points to a smaller 
number of radio-bright early type stars compared to what would be expected 
on the basis of a typical stellar IMF alone. For instance, according to the Kroupa 
(2001) IMF, roughly 50\% of all stars have a spectral type between K0 and M5.
Could this be due to an IMF inherently deficient in higher-mass stars? Hsu et al.\
(2013) find that the pre-main sequence population of the L1641 
dark cloud in Orion, has a deficit of higher mass stars when compared to a 
compact region like the Orion Nebula Cluster, and suggest that the environment
may play a role in determining the high-mass end of the IMF. We note, however, 
that Erickson et al.\ (2011) constructed an IMF for an extinction-limited sample of 
123 YSOs in L1688 and found it to be consistent with that of  field star for M $>$ 
0.2 \Msun. Alternatively, our finding might indicate that the fraction of radio-bright young 
stars increases for later type stars, in agreement with the trend noticed for 
more evolved stars by Berger et al. (2010).  On the other hand, the detected 
F/A/B stars are on average significantly brighter at radio wavelengths than their 
M and K counterparts, confirming that (for earlier type stars), there is a good 
correlation between the bolometric and radio luminosities.

It is also possible that the lower fraction of radio-bright early type stars could be the 
result of an observational bias. The problem is that it is very difficult to identify
YSOs with spectral types G, F, A and B, unless they are 
actively accreting, as inferred from certain lines strongly in emission,
like H$\alpha$ 6563{\AA } and He I 5875,6676{\AA}, and
forbidden lines like [OI] or [SII], or have infrared excess emission originating in a
circumstellar disk.While for T Tauri stars we have well defined observables 
to identify the full YSO population, for the earlier type counterparts 
this is not the case.
A strong Li~I 6707{\AA } absorption line, 
in excess of what is observed in young main sequence stars like the 
Pleiades (Brice\~no et al.\ 1997, 1998) does not imply youth in
G and earlier type stars, because they do not deplete Li~I during their
pre-main sequence phase.
The strength of the Na~I (8183,8195\AA) doublet, which allows to distinguish 
low surface gravity stars still contracting toward the main sequence
(e.g.\ Martin et al.\ 1996; Slesnick et al.\ 2006; Downes et al.\ 2008; Lodieu 
et al.\ 2011; Schlieder et al.\ 2012), breaks down for spectral types earlier than 
about M0. Strong X-ray emission, characteristic of late F and G through M-type 
young stars, unless combined with other criteria, can suffer from significant 
contamination by young main sequence stars with ages up to $\sim 100$ Myr 
(Brice\~no et al.\ 1997).
So far, the best way to determine the full membership of G through B-type YSOs
seems to be selecting as members objects which satisfy as primary criteria
radial velocities and proper motions (if possible, e.g.\ Hsun et al.\ 2013), 
combine these with other
characteristics like X-ray emission, emission lines, infrared excesses,
and place them in a H-R diagram,
which requires a reasonable knowledge of the distance to each 
and every object, or assumption of a common distance to a group or cluster.
If, as suggested by the trend observed in Figure 5, Class III
sources are indeed, on average, brighter radio sources than Class II YSOs,
the fact that the early-type sample may be biased toward the later
could be affecting our result.

\subsection{The radio -- X-ray relation}

G\"udel \& Benz (1993) and Benz \& G\"udel (1994) showed that the radio and X-ray 
emissions of magnetically active stars are correlated by a relation of the form:

\begin{displaymath}
 \frac{L_X}{L_R}=\kappa\cdot10^{15.5\pm1}\quad{\rm [Hz]}.
\end{displaymath}

\noindent
where $\kappa$ is unity for dMe and dK stars, BY Dra-type binaries and RS
CVn binaries with two subgiants. For classical RS CVn binaries, Algol
systems, FK Com stars and Post T Tauri stars $\kappa\approx0.17$. The
radio observations used to establish this relation were obtained mainly at 4.8 GHz 
and, in some cases, at 8.4 GHz. 

Of the 55 young stars in our sample, 35 have known X-ray counterparts, and can be used to 
study the $L_X/L_R$ relation for young stellar objects in a statistically significant fashion. Note
that we corrected all X-ray luminosities to the distance of 120 pc adopted in this work. In Figure 
\ref{fig:GB}, we show the X-ray luminosities of the young stellar objects as a function of their 
radio luminosities at both radio frequencies observed in this work. In agreement with the 
results obtained by Gagn\'e et al.\ (2004) and Forbrich et al.\ (2010), $L_X/L_R$ $\leq10^{15.5}$ 
for our sample. Indeed, a relation $L_X/L_R\approx10^{14\pm1}$ provides a good match to the 
distribution of points in this plot. This is equivalent, in terms of the G\"udel-Benz relation, to 
$\kappa$ = 0.03 for young stellar objects. We argue that this results provides an extension 
of the relation valid for YSOs.

\section{Comments on some individual sources}

\subsection{VLA 1623}

A total of four radio sources associated with the Class~0 source VLA~1623 have been reported 
in the literature. Bontemps et al.\ (1997; see their Figures 1 and 2) reported on the detection of three 
roughly aligned objects that they called A, B, and C (from east to west). At higher angular resolution, 
however, their source A breaks down into two sub-condensation. They interpret the  easternmost 
(and weakest) of these two sub-condensations as the protostar VLA~1623 itself, and the western 
sub-condensation, as well as the sources B and C, as knots along a jet driven by VLA~1623. In
more recent publications, the two sub-condensations within source A of Bontemps et al.\ (1995)
have usually been referred to as VLA~1623A and VLA~1623B (e.g.\ Ward-Thompson et al.\ 2011; 
Murillo \& Lai 2013), while source B of Bontemps et al.\ (1995) has been renamed VLA~1623W;
source C has, to our knowledge, never been detected again. While there is general agreement that 
VLA~1623A is a protostar, the nature of VLA~1623B and VLA~1623W is still debated.\footnote{Note,
however, that both have measurable proper motions similar to those of other sources in Ophiuchus
(J.L.\ Rivera et al., in prep). Thus, they are associated with that region: neither is an related background 
or foreground objects.} 
Looney et al.\  (2000) obtained dust observations at $\lambda$ = 2.7 mm and concluded that VLA~1623B 
was, most likely, a stellar companion of VLA~1623A. A similar conclusion was reached by Murillo \& Lai (2013) 
based on Sub-Millimeter Array (SMA) observations. The latter authors detected VLA~1623W as a 
bright point source in the MIPS 24 micron data. This strongly suggest that it might 
also be a protostellar source, making VLA~1623 a triple system. However, some properties of the
radio emission from VLA~1623B are more easily explained if it is interpreted as a knot along the jet 
(Ward-Thompson et al.\ 2011). 

VLA~1623 A, B, and W are detected in our observations as GBS-VLA J162626.39-242430.9, 
GBS-VLA J162626.31-242430.3, and GBS-VLA J162625.62-242429.2, respectively. Source C
of Bontemps et al.\ (1997), however, is not seen in our data. Both VLA~1623A and VLA~1623B 
have positive spectral indices of order 0.8, more typical of protostellar sources than of outflow
features. VLA~1623W, on the other hand, has a spectral index (--0.2 $\pm$ 0.6) consistent with 
the optically thin free-free emission expected from an outflow knot (e.g.\ Pech et al.\ 2010). 
Interestingly, however, the separations between VLA~1623A and VLA~1623B (1.2 arcsec), 
and between VLA~1623A and VLA~1623W (10.5 arcsec) have not changed appreciably 
during the $\sim$ 15 years separating the observations reported by Bontemps et al.\ (1997) 
and those described here. This sets an upper limit of about 4 km s$^{-1}$ on their relative motion. 
While this would be consistent with the expected orbital motion in a solar mass multiple
system, it is harder to reconcile with sources B and W being outflow knots. 

\subsection{Radio detections at the stellar-brown dwarf boundary}

We detect radio emission from two brown dwarf candidates: GBS-VLA J162722.96-242236.6
is associated with a brown dwarf reported and documented by Marsh et al. (2010),
while GBS-VLA J162715.70-243845.6 is a brown dwarf candidate (Alves de Oliveira et 
al.\ 2010) located in the southernmost component in the triple system WL 20. We note that the
association of the radio source with the brown dwarf candidate in WL 20 (rather than 
with the higher mass members of the system) is secure since the angular offset between the
position of the radio source and the nominal position of the brown dwarf candidate is 
0\rlap{$''$}.14, while its separations from the other stars in the system are 2\rlap{$''$}.42 
and 3\rlap{$''$}.31. We also detected radio emission from two M stars near the brown dwarf 
boundary: GBS-VLA J162556.09-243015.3 is an M5 young stellar object of Class III, while 
GBS-VLA J162759.95-244819.5 is an M4.75 weak line T Tauri star.

We will discuss these cases in more details in a forthcoming dedicated publication, but would
like to note here that in three of these four objects near the brown dwarf boundary, the radio 
emission shows clear indications (high levels of variability and/or a negative spectral
index) of being non-thermal. Only in the case of GBS-VLA J162556.09-243015.3
is the emission more likely thermal (free-free) since the spectral index is around
zero and the radio flux quite steady. Our detections show that young stellar objects near
or beyond the brown dwarf boundary can be detectable radio emitters even at a
distance of more than 100 pc. Like their older counterparts (e.g.\ McLean et al.\ 2011; Ravi 
et al.\ 2011), they appear to have a large radio to bolometric luminosity ratio.

\subsection{Non-thermal Radio Emission from Class I Objects}

Protostellar objects (of Class 0 and I) are expected to have strong winds
producing thermal bremsstrahlung (free-free) emission that is optically thick 
at least in the dense region immediately surrounding the protostar (e.g.\
Rodr\'iguez 1999). In this situation, even if the protostar itself emitted 
non-thermal (gyrosynchrotron) radiation, it ought to be absorbed by the
ionized wind and should not reach the observer. It is noteworthy, however,
that a small number of Class~I sources have been found to be non-thermal 
emitters. Perhaps the most robust cases are IRS5 in Corona Australis (Feigelson 
et al.\ 1998; Deller,
Forbrich \& Loinard 2013), EC 95 in Serpens (Dzib et al.\ 2010), and (more
marginally) YLW~15 in Ophiuchus (Forbrich et al.\ 2007). The detectability 
of non-thermal emission in these objects might be due to a favorable
geometry (e.g.\ when the protostar is seen nearly pole-on or nearly edge-on,
the free-free opacity might be reduced; Forbrich et al.\ 1997), or to tidal clearing
of circumstellar material in a tight binary system (Dzib et al.\ 2010). 

We find two possible non-thermal Class~I sources in our sample. On the
one hand, GBS-VLA J162726.90-244050.8 corresponds to YLW 15 which, as
we just mentioned, is one of the known candidates non-thermal Class~I
sources. In our observations, its spectral index is negative (suggesting a
non-thermal process) but its flux is almost constant. Forbrich et al.\ (2007) 
marginally detected one of the components in YLW~15 during VLBI observations
at 8.4 GHz, at a level of 145 $\mu$Jy. An independent, more robust, VLBI 
detection would be necessary to confirm the non-thermal nature of the 
source. The other target of interest in this context is GBS-VLA J163200.97-245643.3 
which is associated with the Class~I source WLY 2-67. While it has a positive spectral 
index, it exhibits significant variability and it is found to be significantly circularly
polarized (11 and 25 \% at 4.5 and 7.5 GHz, respectively). This provides a clear
indication that the radio emission is non-thermal. A VLBI detection should be
attempted of this target, as it would provide a direct and independent confirmation 
of the non-thermal origin of the emission.

\subsection{{\bf GBS-VLA} J162700.00-242640.3: a new calibrator for Ophiuchus}

One of the known extragalactic targets detected here (GBS-VLA J162700.00-242640.3) 
is found to have a flux of order 71 mJy both at 4.5 and 7.5 GHz, to exhibit very little 
variability ($\sim$ 15\%; Table \ref{tab:sources}) and to be unresolved in
our VLA data. It has been detected in a number of previous radio observations 
(Table \ref{tab:counterparts}) as well as in high-resolution Ka-band observations 
(at $\nu$ = 32.5 GHz) that we obtained in 2011 in the BnA configuration of the VLA. 
The spectral energy distribution constructed from all available radio data (Figure
\ref{fig:calibrator}) shows that it is a flat spectrum source. We have observed it with
the Very Long Baseline Array (VLBA) at $\nu$ = 8.4 GHz, and detected it as an 
unresolved 70 mJy source. Since GBS-VLA J162700.00-242640.3 is located in the
direction of the Ophiuchus core (Figure \ref{fig:smap}), it ought to be used as main
gain calibrator for any future interferometric (conventional or long baseline) 
observations of the Ophiuchus complex.

\section{Conclusions and perspectives}

In this paper, we have reported on radio observations of the Ophiuchus complex
that largely surpass all such previous observations thanks to a combination of 
high sensitivity, good angular resolution, and large field of view. A total of 189
sources were detected, 56 of them associated with known young stellar sources, 
and 4 with known extragalactic objects; the other 129 remain unclassified, but 
most of them are certainly extragalactic background sources. Most of the young 
stars detected at radio wavelengths have spectral types K or M, but we also detect
2 brown dwarf candidates.

Interestingly, at least half of these young stars are non-thermal (gyrosynchrotron) 
sources, with active magnetized coronas. These sources are excellent targets for
future astrometric observations with VLBI instruments (similar to, but much more
extensive than those reported by Loinard et al.\ 2008) that would enable an  accurate 
determination of the distance, kinematics, and internal structure of the Ophiuchus 
region. Such observations would be greatly aided by the detection, reported here, 
of an adequate gain calibrator (GBS-VLA J162700.00-242640.3) located toward the
Ophiuchus core.

\acknowledgments
L.L.\ is grateful to the von Humboldt Stiftung for financial support. S.D., L.L., L.F.R.,
G.N.O., G.P., and J.L.R.\ acknowledge the financial support of DGAPA, UNAM and 
CONACyT, M\'exico. The National Radio Astronomy Observatory 
is operated by Associated Universities Inc. under cooperative agreement with the National 
Science Foundation. CASA is developed by an international consortium of 
scientists based at the National Radio Astronomical Observatory (NRAO), the European 
Southern Observatory (ESO), the National Astronomical Observatory of 
Japan (NAOJ), the CSIRO Australia Telescope National Facility (CSIRO/ATNF), 
and the Netherlands Institute for Radio Astronomy (ASTRON) under the 
guidance of NRAO. This research has made use of the SIMBAD database,
operated at CDS, Strasbourg, France

\begin{deluxetable}{ccrcrc}
\tabletypesize{\scriptsize}
\tablewidth{0pt}
\tablecolumns{6}
\tablecaption{Radio Sources Detected in Ophiuchus \label{tab:sources}}
\tablehead{           & \multicolumn{4}{c}{Flux Properties}	&Spectral\\
\colhead{GB-VLA Name} & \colhead{$f_{4.5}$(mJy)} &\colhead{Var.\,(\%)} & \colhead{$f_{7.5}$(mJy)} & \colhead{Var.\,(\%)}&\colhead{Index}\\
}\startdata
J162540.94-244147.2&(4.39$\pm$0.43$\pm$0.22)$\times10^{-1}$&16.2$\pm$11.9&(2.94$\pm$0.42$\pm$0.15)$\times10^{-1}$&$>$56.2$\pm$12.3&-0.81$\pm$0.38\\
J162542.48-242143.7&(3.79$\pm$0.32$\pm$0.19)$\times10^{-1}$&4.3$\pm$13.7&--&--&--\\
J162547.68-243735.7&(4.19$\pm$0.21$\pm$0.21)$\times10^{-1}$&28.5$\pm$8.6&(4.06$\pm$0.28$\pm$0.20)$\times10^{-1}$&51.3$\pm$15.0&-0.06$\pm$0.22\\
J162548.96-244049.5&(6.49$\pm$0.34$\pm$0.32)$\times10^{-1}$&41.7$\pm$5.4&(2.92$\pm$0.32$\pm$0.15)$\times10^{-1}$&$>$66.9$\pm$4.3&-1.61$\pm$0.28\\
J162550.51-243914.7&(3.57$\pm$0.25$\pm$0.18)$\times10^{-1}$&78.2$\pm$5.3&(2.84$\pm$0.18$\pm$0.14)$\times10^{-1}$&41.4$\pm$13.1&-0.46$\pm$0.24\\
J162550.84-243719.6&(8.00$\pm$1.00$\pm$0.40)$\times10^{-2}$&$>$43.6$\pm$4.7&$<$0.06&--&$<$-0.58$\pm$0.14\\
J162551.91-242917.0&(1.71$\pm$0.35$\pm$0.09)$\times10^{-1}$&21.4$\pm$19.7&(1.15$\pm$0.24$\pm$0.06)$\times10^{-1}$&$>$17.4$\pm$37.5&-0.8$\pm$0.61\\
J162556.09-243015.3&(6.72$\pm$0.15$\pm$0.34)$\times10^{-1}$&31.4$\pm$3.3&(6.76$\pm$0.24$\pm$0.34)$\times10^{-1}$&10.5$\pm$4.5&0.01$\pm$0.17\\
J162557.51-243032.1&(4.61$\pm$0.17$\pm$0.23)$\times10^{-1}$&55.8$\pm$3.3&(4.07$\pm$0.17$\pm$0.20)$\times10^{-1}$&56.1$\pm$4.5&-0.25$\pm$0.18\\
J162603.01-242336.4&(2.67$\pm$0.09$\pm$0.13)$\times10^{+0}$&81.7$\pm$0.6&(2.52$\pm$0.09$\pm$0.13)$\times10^{+0}$&84.9$\pm$1.0&-0.12$\pm$0.18\\
J162605.29-243436.6&$<$0.05&--&(1.05$\pm$0.15$\pm$0.05)$\times10^{-1}$&33.9$\pm$27.2&$>$1.34$\pm$0.14\\
J162607.63-242741.7&(2.10$\pm$0.21$\pm$0.10)$\times10^{-1}$&$>$79.5$\pm$1.0&(1.01$\pm$0.21$\pm$0.05)$\times10^{-1}$&$>$67.0$\pm$3.6&-1.48$\pm$0.49\\
J162607.63-243648.9&(5.83$\pm$0.17$\pm$0.29)$\times10^{-1}$&37.1$\pm$3.1&(3.67$\pm$0.14$\pm$0.18)$\times10^{-1}$&39.4$\pm$7.4&-0.94$\pm$0.17\\
J162608.04-242523.1&(1.03$\pm$0.14$\pm$0.05)$\times10^{-1}$&37.9$\pm$15.6&$<$0.05&--&$<$-1.3$\pm$0.14\\
J162610.32-242054.9&(1.00$\pm$0.12$\pm$0.05)$\times10^{-1}$&$>$56.1$\pm$5.6&(1.60$\pm$0.22$\pm$0.08)$\times10^{-1}$&53.4$\pm$16.7&0.95$\pm$0.40\\
J162610.55-242853.3&(2.09$\pm$0.15$\pm$0.10)$\times10^{-1}$&24.2$\pm$10.0&(1.68$\pm$0.19$\pm$0.08)$\times10^{-1}$&43.0$\pm$17.0&-0.44$\pm$0.31\\
J162611.08-242907.4&(4.51$\pm$0.10$\pm$0.23)$\times10^{-1}$&16.2$\pm$4.4&(2.98$\pm$0.16$\pm$0.15)$\times10^{-1}$&42.4$\pm$6.5&-0.84$\pm$0.18\\
J162611.79-243716.4&(1.66$\pm$0.12$\pm$0.08)$\times10^{-1}$&20.9$\pm$15.5&(1.11$\pm$0.12$\pm$0.06)$\times10^{-1}$&30.9$\pm$17.5&-0.81$\pm$0.30\\
J162615.67-243421.2&(9.17$\pm$0.07$\pm$0.46)$\times10^{+0}$&--&(6.23$\pm$0.12$\pm$0.31)$\times10^{+0}$&--&Extended\\
J162616.84-242223.5&(3.37$\pm$0.15$\pm$0.17)$\times10^{-1}$&$>$75.9$\pm$0.8&(3.60$\pm$0.24$\pm$0.18)$\times10^{-1}$&$>$83.1$\pm$0.3&0.13$\pm$0.22\\
J162617.66-244014.5&(4.65$\pm$0.26$\pm$0.23)$\times10^{-1}$&46.3$\pm$5.0&(3.31$\pm$0.25$\pm$0.17)$\times10^{-1}$&30.9$\pm$11.5&-0.69$\pm$0.24\\
J162620.56-243523.4&(7.30$\pm$1.50$\pm$0.36)$\times10^{-2}$&$>$40.2$\pm$23.0&$<$0.05&--&$<$-0.61$\pm$0.14\\
J162621.72-242250.7&(2.38$\pm$0.17$\pm$0.12)$\times10^{-1}$&28.8$\pm$11.2&(3.04$\pm$0.29$\pm$0.15)$\times10^{-1}$&$>$13.2$\pm$14.2&0.49$\pm$0.28\\
J162622.38-242253.3&(2.02$\pm$0.06$\pm$0.10)$\times10^{+0}$&68.6$\pm$1.6&(1.42$\pm$0.07$\pm$0.07)$\times10^{+0}$&85.1$\pm$1.4&-0.71$\pm$0.18\\
J162623.40-243223.6&(1.25$\pm$0.15$\pm$0.06)$\times10^{-1}$&$>$37.9$\pm$11.4&$<$0.06&--&$<$-1.48$\pm$0.14\\
J162623.57-242439.6&(1.24$\pm$0.17$\pm$0.06)$\times10^{-1}$&33.5$\pm$16.2&(1.15$\pm$0.08$\pm$0.06)$\times10^{-1}$&$>$38.2$\pm$6.0&-0.15$\pm$0.34\\
J162624.08-241613.5&(1.92$\pm$0.15$\pm$0.10)$\times10^{-1}$&22.3$\pm$9.6&(3.44$\pm$0.28$\pm$0.17)$\times10^{-1}$&46.6$\pm$6.5&1.18$\pm$0.27\\
J162625.62-242429.2&(2.18$\pm$0.14$\pm$0.11)$\times10^{-1}$&17.6$\pm$15.2&(1.98$\pm$0.23$\pm$0.10)$\times10^{-1}$&$>$47.2$\pm$6.3&-0.19$\pm$0.30\\
J162626.03-244923.7&(1.73$\pm$0.05$\pm$0.09)$\times10^{+1}$&--&--&--&Extended\\
J162626.31-242430.2&(1.89$\pm$0.34$\pm$0.09)$\times10^{-1}$&63.1$\pm$11.0&(1.89$\pm$0.34$\pm$0.09)$\times10^{-1}$&44.7$\pm$14.6&-0.0$\pm$0.53\\
J162626.39-242430.9&(8.70$\pm$3.00$\pm$0.43)$\times10^{-2}$&--\tablenotemark{a}&(1.25$\pm$0.25$\pm$0.06)$\times10^{-1}$&--\tablenotemark{a}&0.73$\pm$0.82\\
J162629.63-242317.2&(2.28$\pm$0.14$\pm$0.11)$\times10^{-1}$&43.3$\pm$10.3&(1.24$\pm$0.18$\pm$0.06)$\times10^{-1}$&$>$64.2$\pm$9.4&-1.23$\pm$0.35\\
J162629.67-241905.8&(2.67$\pm$0.17$\pm$0.13)$\times10^{-1}$&57.4$\pm$6.0&(2.27$\pm$0.21$\pm$0.11)$\times10^{-1}$&69.9$\pm$5.2&-0.33$\pm$0.27\\
J162630.15-243132.4&(1.08$\pm$0.20$\pm$0.05)$\times10^{-1}$&$>$59.2$\pm$8.6&$<$0.05&--&$<$-1.4$\pm$0.14\\
J162630.59-242023.0&(9.80$\pm$1.30$\pm$0.49)$\times10^{-2}$&36.9$\pm$17.1&(6.40$\pm$1.70$\pm$0.32)$\times10^{-2}$&--\tablenotemark{a}&-0.86$\pm$0.62\\
J162631.28-241832.9&(2.42$\pm$0.26$\pm$0.12)$\times10^{-1}$&43.8$\pm$10.6&(1.39$\pm$0.15$\pm$0.07)$\times10^{-1}$&8.8$\pm$30.6&-1.12$\pm$0.34\\
J162631.34-243341.8&(2.22$\pm$0.56$\pm$0.11)$\times10^{-1}$&37.1$\pm$19.4&(1.66$\pm$0.26$\pm$0.08)$\times10^{-1}$&$>$67.9$\pm$11.9&-0.59$\pm$0.62\\
J162632.78-241627.5&(9.30$\pm$1.50$\pm$0.46)$\times10^{-2}$&38.8$\pm$18.7&(1.42$\pm$0.24$\pm$0.07)$\times10^{-1}$&32.1$\pm$18.4&0.86$\pm$0.49\\
J162633.16-245246.7&(8.03$\pm$0.27$\pm$0.40)$\times10^{-1}$&30.7$\pm$5.5&(9.18$\pm$0.43$\pm$0.46)$\times10^{-1}$&49.3$\pm$5.3&0.27$\pm$0.18\\
J162633.48-241215.9&(2.12$\pm$0.03$\pm$0.11)$\times10^{+0}$&17.4$\pm$1.5&(9.27$\pm$0.32$\pm$0.46)$\times10^{-1}$&38.1$\pm$5.3&-1.67$\pm$0.16\\
J162634.17-242328.4&(7.98$\pm$0.25$\pm$0.40)$\times10^{+0}$&12.0$\pm$3.4&(7.07$\pm$0.34$\pm$0.35)$\times10^{+0}$&23.6$\pm$5.3&-0.24$\pm$0.18\\
J162634.47-241656.8&(9.20$\pm$0.70$\pm$0.46)$\times10^{-2}$&$>$41.0$\pm$3.0&$<$0.04&--&$<$-1.44$\pm$0.14\\
J162634.95-242655.3&(1.97$\pm$0.19$\pm$0.10)$\times10^{-1}$&30.1$\pm$12.9&(1.00$\pm$0.13$\pm$0.05)$\times10^{-1}$&40.5$\pm$13.9&-1.37$\pm$0.36\\
J162635.33-242405.2&(6.50$\pm$0.38$\pm$0.33)$\times10^{-1}$&9.3$\pm$9.3&(3.29$\pm$0.33$\pm$0.16)$\times10^{-1}$&20.3$\pm$18.6&-1.38$\pm$0.27\\
J162636.96-243755.9&(4.00$\pm$0.26$\pm$0.20)$\times10^{-1}$&12.8$\pm$10.7&(1.75$\pm$0.24$\pm$0.09)$\times10^{-1}$&25.4$\pm$15.5&-1.67$\pm$0.34\\
J162637.27-244553.9&(2.91$\pm$0.33$\pm$0.15)$\times10^{-1}$&52.8$\pm$8.4&(1.81$\pm$0.26$\pm$0.09)$\times10^{-1}$&$>$78.0$\pm$3.0&-0.96$\pm$0.40\\
J162638.99-244529.0&(3.78$\pm$0.19$\pm$0.19)$\times10^{-1}$&12.8$\pm$10.5&(2.29$\pm$0.50$\pm$0.11)$\times10^{-1}$&$>$24.6$\pm$26.4&-1.01$\pm$0.47\\
J162639.00-243052.8&(4.05$\pm$0.27$\pm$0.20)$\times10^{-1}$&43.6$\pm$9.1&(2.37$\pm$0.19$\pm$0.12)$\times10^{-1}$&21.4$\pm$14.8&-1.08$\pm$0.25\\
J162639.70-241609.7&(1.19$\pm$0.16$\pm$0.06)$\times10^{-1}$&15.3$\pm$31.0&(6.60$\pm$1.40$\pm$0.33)$\times10^{-2}$&--\tablenotemark{a}&-1.19$\pm$0.53\\
J162641.11-245855.8&(6.45$\pm$0.26$\pm$0.32)$\times10^{-1}$&67.8$\pm$3.3&(8.25$\pm$0.27$\pm$0.41)$\times10^{-1}$&81.0$\pm$2.3&0.5$\pm$0.18\\
J162642.44-242626.1&(2.66$\pm$0.06$\pm$0.13)$\times10^{+0}$&88.0$\pm$0.5&(2.48$\pm$0.10$\pm$0.12)$\times10^{+0}$&92.3$\pm$1.2&-0.14$\pm$0.17\\
J162642.53-244628.3&(9.57$\pm$0.17$\pm$0.48)$\times10^{-1}$&51.3$\pm$3.2&(8.31$\pm$0.43$\pm$0.42)$\times10^{-1}$&35.4$\pm$4.5&-0.29$\pm$0.18\\
J162643.76-241633.4&(1.31$\pm$0.04$\pm$0.07)$\times10^{+0}$&45.6$\pm$1.8&(1.35$\pm$0.04$\pm$0.07)$\times10^{+0}$&46.0$\pm$1.8&0.07$\pm$0.17\\
J162646.36-242002.0&(5.53$\pm$0.21$\pm$0.28)$\times10^{-1}$&15.3$\pm$6.4&(7.68$\pm$0.20$\pm$0.38)$\times10^{-1}$&21.6$\pm$5.7&0.66$\pm$0.17\\
J162647.17-245157.4&(5.91$\pm$0.25$\pm$0.30)$\times10^{-1}$&67.5$\pm$5.3&(3.51$\pm$0.84$\pm$0.18)$\times10^{-1}$&$>$39.9$\pm$14.0&-1.05$\pm$0.51\\
J162647.23-243620.3&(4.18$\pm$0.15$\pm$0.21)$\times10^{-1}$&21.5$\pm$6.9&(4.87$\pm$0.24$\pm$0.24)$\times10^{-1}$&22.0$\pm$7.9&0.31$\pm$0.19\\
J162647.32-245852.6&(1.50$\pm$0.25$\pm$0.07)$\times10^{-1}$&$>$12.2$\pm$16.1&$<$0.13&--&$<$-0.29$\pm$0.14\\
J162647.57-245754.8&(5.75$\pm$0.62$\pm$0.29)$\times10^{-1}$&16.6$\pm$13.3&(3.08$\pm$0.28$\pm$0.15)$\times10^{-1}$&37.4$\pm$9.8&-1.26$\pm$0.32\\
J162649.23-242003.3&(9.95$\pm$0.27$\pm$0.50)$\times10^{-1}$&82.0$\pm$2.0&(7.44$\pm$0.40$\pm$0.37)$\times10^{-1}$&85.0$\pm$2.3&-0.59$\pm$0.19\\
J162649.90-245617.5&(2.02$\pm$0.15$\pm$0.10)$\times10^{-1}$&35.7$\pm$11.6&(1.87$\pm$0.18$\pm$0.09)$\times10^{-1}$&33.3$\pm$21.8&-0.16$\pm$0.28\\
J162651.69-241441.5&(2.10$\pm$0.16$\pm$0.10)$\times10^{-1}$&$>$78.9$\pm$0.9&(1.47$\pm$0.17$\pm$0.07)$\times10^{-1}$&$>$71.2$\pm$2.3&-0.72$\pm$0.31\\
J162653.38-241105.8&(2.83$\pm$0.20$\pm$0.14)$\times10^{-1}$&55.0$\pm$6.8&(2.76$\pm$0.16$\pm$0.14)$\times10^{-1}$&33.1$\pm$7.6&-0.05$\pm$0.23\\
J162657.84-244201.6&(1.74$\pm$0.18$\pm$0.09)$\times10^{-1}$&35.9$\pm$11.7&(1.61$\pm$0.16$\pm$0.08)$\times10^{-1}$&50.6$\pm$11.6&-0.16$\pm$0.32\\
J162658.25-243738.5&(1.12$\pm$0.11$\pm$0.06)$\times10^{-1}$&57.3$\pm$11.1&(5.40$\pm$1.30$\pm$0.27)$\times10^{-2}$&--\tablenotemark{a}&-1.47$\pm$0.54\\
J162658.38-242130.5&(2.04$\pm$0.17$\pm$0.10)$\times10^{-1}$&$>$79.4$\pm$0.8&(1.62$\pm$0.28$\pm$0.08)$\times10^{-1}$&$>$85.8$\pm$0.7&-0.47$\pm$0.41\\
J162659.16-243458.9&(5.51$\pm$0.16$\pm$0.28)$\times10^{-1}$&10.3$\pm$6.4&(7.02$\pm$0.24$\pm$0.35)$\times10^{-1}$&12.5$\pm$4.9&0.49$\pm$0.17\\
J162659.96-243639.5&(1.42$\pm$0.20$\pm$0.07)$\times10^{-1}$&34.5$\pm$9.0&$<$0.05&--&$<$-1.95$\pm$0.14\\
J162659.99-241910.0&$<$0.10&--&(5.35$\pm$0.59$\pm$0.27)$\times10^{-1}$&$>$84.5$\pm$1.0&$>$3.39$\pm$0.14\\
J162700.00-242640.3&(7.11$\pm$0.15$\pm$0.36)$\times10^{+1}$&12.6$\pm$1.8&(7.09$\pm$0.38$\pm$0.35)$\times10^{+1}$&17.3$\pm$2.7&-0.01$\pm$0.18\\
J162700.02-243537.6&(2.45$\pm$0.13$\pm$0.12)$\times10^{-1}$&24.0$\pm$12.4&(1.91$\pm$0.13$\pm$0.10)$\times10^{-1}$&46.5$\pm$11.9&-0.5$\pm$0.23\\
J162702.11-243842.5&(2.20$\pm$0.17$\pm$0.11)$\times10^{-1}$&$>$55.5$\pm$4.3&(2.24$\pm$0.16$\pm$0.11)$\times10^{-1}$&68.6$\pm$7.3&0.04$\pm$0.26\\
J162702.15-241927.8&(5.48$\pm$0.19$\pm$0.27)$\times10^{-1}$&6.4$\pm$8.7&(3.09$\pm$0.16$\pm$0.15)$\times10^{-1}$&19.3$\pm$11.8&-1.16$\pm$0.19\\
J162702.33-243727.3&(1.02$\pm$0.16$\pm$0.05)$\times10^{-1}$&$>$43.0$\pm$9.7&(1.21$\pm$0.14$\pm$0.06)$\times10^{-1}$&$>$28.4$\pm$10.2&0.35$\pm$0.42\\
J162702.36-242724.8&(3.62$\pm$0.19$\pm$0.18)$\times10^{-1}$&$>$4.5$\pm$10.5&(2.32$\pm$0.25$\pm$0.12)$\times10^{-1}$&$>$39.7$\pm$5.3&-0.9$\pm$0.28\\
J162705.16-242007.8&(2.54$\pm$0.22$\pm$0.13)$\times10^{-1}$&27.9$\pm$13.7&(1.41$\pm$0.15$\pm$0.07)$\times10^{-1}$&$>$43.6$\pm$6.7&-1.19$\pm$0.31\\
J162705.25-243629.8&(1.46$\pm$0.30$\pm$0.07)$\times10^{-1}$&5.6$\pm$30.7&(1.33$\pm$0.14$\pm$0.07)$\times10^{-1}$&29.6$\pm$15.1&-0.19$\pm$0.49\\
J162705.96-242618.9&(3.85$\pm$0.30$\pm$0.19)$\times10^{-1}$&$>$76.8$\pm$1.5&(2.89$\pm$0.21$\pm$0.14)$\times10^{-1}$&$>$6.4$\pm$7.2&-0.58$\pm$0.26\\
J162709.41-243719.0&(2.51$\pm$0.15$\pm$0.13)$\times10^{-1}$&30.9$\pm$12.1&(3.57$\pm$0.16$\pm$0.18)$\times10^{-1}$&24.2$\pm$8.9&0.71$\pm$0.21\\
J162711.29-243722.5&(1.45$\pm$0.12$\pm$0.07)$\times10^{-1}$&20.6$\pm$13.9&(8.00$\pm$1.00$\pm$0.40)$\times10^{-2}$&$>$44.8$\pm$13.8&-1.2$\pm$0.33\\
J162713.06-241817.0&(3.00$\pm$0.18$\pm$0.15)$\times10^{-1}$&41.2$\pm$6.0&(1.79$\pm$0.18$\pm$0.09)$\times10^{-1}$&21.3$\pm$14.4&-1.04$\pm$0.28\\
J162713.62-242226.1&(1.31$\pm$0.20$\pm$0.07)$\times10^{-1}$&42.7$\pm$24.8&(1.08$\pm$0.26$\pm$0.05)$\times10^{-1}$&$>$31.8$\pm$24.8&-0.39$\pm$0.59\\
J162714.71-243919.7&(1.71$\pm$0.18$\pm$0.09)$\times10^{-1}$&$>$43.6$\pm$4.7&$<$0.07&--&$<$-1.8$\pm$0.14\\
J162715.69-243845.7&(2.20$\pm$0.16$\pm$0.11)$\times10^{-1}$&24.4$\pm$13.2&(2.36$\pm$0.28$\pm$0.12)$\times10^{-1}$&35.1$\pm$10.7&0.14$\pm$0.32\\
J162717.38-243616.7&(1.20$\pm$0.14$\pm$0.06)$\times10^{-1}$&22.5$\pm$18.5&$<$0.05&--&$<$-1.61$\pm$0.14\\
J162718.17-242852.9&(3.78$\pm$0.07$\pm$0.19)$\times10^{+0}$&68.7$\pm$1.0&(3.14$\pm$0.09$\pm$0.16)$\times10^{+0}$&65.2$\pm$1.6&-0.38$\pm$0.16\\
J162718.25-243334.8&(2.99$\pm$0.28$\pm$0.15)$\times10^{-1}$&$>$45.8$\pm$3.0&(1.14$\pm$0.14$\pm$0.06)$\times10^{-1}$&$>$53.1$\pm$17.7&-1.95$\pm$0.34\\
J162719.34-243130.4&(1.21$\pm$0.18$\pm$0.06)$\times10^{-1}$&$>$28.6$\pm$21.1&(8.50$\pm$1.20$\pm$0.43)$\times10^{-2}$&--\tablenotemark{a}&-0.71$\pm$0.44\\
J162719.50-244140.3&(1.60$\pm$0.37$\pm$0.08)$\times10^{-1}$&$>$42.4$\pm$12.0&(8.70$\pm$1.20$\pm$0.43)$\times10^{-2}$&$>$56.6$\pm$5.5&-1.23$\pm$0.56\\
J162721.81-244335.9&(3.66$\pm$0.23$\pm$0.18)$\times10^{-1}$&76.1$\pm$3.3&(3.42$\pm$0.22$\pm$0.17)$\times10^{-1}$&$>$81.8$\pm$0.9&-0.14$\pm$0.23\\
J162721.97-242940.0&(1.47$\pm$0.24$\pm$0.07)$\times10^{-1}$&45.9$\pm$17.9&(1.32$\pm$0.16$\pm$0.07)$\times10^{-1}$&41.1$\pm$13.0&-0.22$\pm$0.43\\
J162722.96-242236.6&(2.09$\pm$0.15$\pm$0.10)$\times10^{-1}$&66.8$\pm$10.6&$<$0.06&--&$<$-2.7$\pm$0.14\\
J162724.19-242929.8&(1.52$\pm$0.17$\pm$0.08)$\times10^{-1}$&$>$57.1$\pm$4.6&(1.52$\pm$0.19$\pm$0.08)$\times10^{-1}$&$>$33.3$\pm$11.1&-0.0$\pm$0.37\\
J162725.40-244659.7&(9.60$\pm$1.20$\pm$0.48)$\times10^{-2}$&$>$26.5$\pm$14.4&--&--&--\\
J162726.90-244050.8&(1.87$\pm$0.06$\pm$0.09)$\times10^{+0}$&6.1$\pm$5.5&(1.53$\pm$0.05$\pm$0.08)$\times10^{+0}$&20.0$\pm$4.7&-0.41$\pm$0.17\\
J162727.36-243116.8&(1.76$\pm$0.20$\pm$0.09)$\times10^{-1}$&58.6$\pm$12.7&(1.26$\pm$0.15$\pm$0.06)$\times10^{-1}$&34.6$\pm$17.9&-0.68$\pm$0.36\\
J162728.00-243933.7&(3.50$\pm$0.22$\pm$0.17)$\times10^{-1}$&22.3$\pm$10.1&(3.96$\pm$0.30$\pm$0.20)$\times10^{-1}$&12.3$\pm$12.8&0.25$\pm$0.24\\
J162729.23-241755.3&(3.90$\pm$0.06$\pm$0.20)$\times10^{+0}$&18.6$\pm$1.3&(2.33$\pm$0.06$\pm$0.12)$\times10^{+0}$&28.9$\pm$2.7&-1.04$\pm$0.15\\
J162730.82-244727.2&(8.44$\pm$0.16$\pm$0.42)$\times10^{-1}$&64.2$\pm$2.2&(4.87$\pm$0.65$\pm$0.24)$\times10^{-1}$&$>$29.7$\pm$7.7&-1.11$\pm$0.31\\
J162731.05-243403.4&(1.64$\pm$0.22$\pm$0.08)$\times10^{-1}$&$>$69.3$\pm$2.2&(3.58$\pm$0.29$\pm$0.18)$\times10^{-1}$&$>$89.0$\pm$0.4&1.58$\pm$0.35\\
J162731.19-242833.8&(1.87$\pm$0.09$\pm$0.09)$\times10^{-1}$&$>$61.3$\pm$1.2&(1.05$\pm$0.05$\pm$0.05)$\times10^{-1}$&51.7$\pm$6.2&-1.17$\pm$0.20\\
J162732.68-243324.5&(1.84$\pm$0.19$\pm$0.09)$\times10^{-1}$&44.7$\pm$13.0&(2.00$\pm$0.15$\pm$0.10)$\times10^{-1}$&61.2$\pm$9.1&0.17$\pm$0.29\\
J162732.76-241401.7&(3.32$\pm$0.19$\pm$0.17)$\times10^{-1}$&58.0$\pm$6.6&--&--&--\\
J162734.55-242020.7&(1.18$\pm$0.03$\pm$0.06)$\times10^{+0}$&29.6$\pm$4.2&(6.72$\pm$0.25$\pm$0.34)$\times10^{-1}$&38.7$\pm$6.0&-1.14$\pm$0.17\\
J162735.13-242624.0&(3.48$\pm$0.12$\pm$0.17)$\times10^{-1}$&17.3$\pm$6.8&(1.72$\pm$0.19$\pm$0.09)$\times10^{-1}$&45.8$\pm$9.4&-1.42$\pm$0.27\\
J162736.00-241402.9&(2.34$\pm$0.24$\pm$0.12)$\times10^{-1}$&$>$58.6$\pm$3.9&--&--&--\\
J162738.20-242630.5&(2.73$\pm$0.13$\pm$0.14)$\times10^{-1}$&12.2$\pm$10.3&(1.98$\pm$0.17$\pm$0.10)$\times10^{-1}$&22.8$\pm$12.7&-0.65$\pm$0.24\\
J162739.42-243915.8&(7.20$\pm$1.50$\pm$0.36)$\times10^{-2}$&44.5$\pm$22.6&(1.25$\pm$0.15$\pm$0.06)$\times10^{-1}$&$>$49.3$\pm$11.0&1.11$\pm$0.51\\
J162740.92-241452.8&(2.68$\pm$0.39$\pm$0.13)$\times10^{+0}$&22.7$\pm$14.7&--&--&--\\
J162741.39-241454.5&(7.38$\pm$0.34$\pm$0.37)$\times10^{-1}$&41.6$\pm$8.2&--&--&--\\
J162741.49-243537.6&(1.43$\pm$0.15$\pm$0.07)$\times10^{-1}$&$>$61.8$\pm$2.6&(1.02$\pm$0.13$\pm$0.05)$\times10^{-1}$&$>$25.7$\pm$13.1&-0.68$\pm$0.36\\
J162742.12-241455.4&(9.70$\pm$1.82$\pm$0.48)$\times10^{-1}$&25.4$\pm$15.4&--&--&--\\
J162745.42-243754.6&(8.20$\pm$1.50$\pm$0.41)$\times10^{-2}$&$>$22.7$\pm$15.1&$<$0.05&--&$<$-0.84$\pm$0.14\\
J162749.85-242540.5&(8.00$\pm$0.10$\pm$0.40)$\times10^{+0}$&32.3$\pm$2.3&(8.51$\pm$0.27$\pm$0.43)$\times10^{+0}$&31.1$\pm$3.1&0.12$\pm$0.16\\
J162751.37-242750.4&(2.86$\pm$0.23$\pm$0.14)$\times10^{-1}$&$>$22.1$\pm$6.8&(1.27$\pm$0.15$\pm$0.06)$\times10^{-1}$&--\tablenotemark{a}&-1.64$\pm$0.32\\
J162751.80-243145.9&(5.98$\pm$1.00$\pm$0.30)$\times10^{-2}$&--\tablenotemark{a}&$<$0.05&--&$<$-0.37$\pm$0.14\\
J162751.89-244630.1&(7.69$\pm$0.42$\pm$0.38)$\times10^{-1}$&$>$95.6$\pm$0.1&(6.61$\pm$0.37$\pm$0.33)$\times10^{-1}$&$>$92.3$\pm$0.4&-0.31$\pm$0.21\\
J162752.08-244050.5&(1.67$\pm$0.05$\pm$0.08)$\times10^{+0}$&76.1$\pm$1.0&(2.21$\pm$0.07$\pm$0.11)$\times10^{+0}$&84.7$\pm$0.6&0.57$\pm$0.17\\
J162752.30-242929.9&(8.70$\pm$1.40$\pm$0.43)$\times10^{-2}$&$>$43.8$\pm$14.8&(1.05$\pm$0.17$\pm$0.05)$\times10^{-1}$&$>$17.1$\pm$16.9&0.38$\pm$0.48\\
J162753.08-242830.3&(2.34$\pm$0.21$\pm$0.12)$\times10^{-1}$&48.9$\pm$11.8&(1.74$\pm$0.29$\pm$0.09)$\times10^{-1}$&$>$45.3$\pm$7.5&-0.6$\pm$0.41\\
J162756.01-244810.7&$<$0.05&--&(1.09$\pm$0.20$\pm$0.05)$\times10^{-1}$&$>$55.8$\pm$15.9&$>$1.53$\pm$0.14\\
J162757.81-244001.9&(3.14$\pm$0.13$\pm$0.16)$\times10^{-1}$&61.3$\pm$4.1&(2.62$\pm$0.13$\pm$0.13)$\times10^{-1}$&64.7$\pm$6.7&-0.37$\pm$0.19\\
J162759.09-242225.9&(2.43$\pm$0.32$\pm$0.12)$\times10^{-1}$&21.5$\pm$15.2&(1.34$\pm$0.22$\pm$0.07)$\times10^{-1}$&$>$25.8$\pm$20.8&-1.2$\pm$0.45\\
J162759.95-244819.5&(1.96$\pm$0.14$\pm$0.10)$\times10^{+0}$&$>$97.0$\pm$0.1&(2.64$\pm$0.21$\pm$0.13)$\times10^{+0}$&$>$95.9$\pm$0.1&0.6$\pm$0.26\\
J162803.39-242124.8&(2.30$\pm$0.22$\pm$0.12)$\times10^{-1}$&39.6$\pm$6.7&$<$0.18&--&$<$-0.5$\pm$0.14\\
J162803.51-242131.2&(1.73$\pm$0.24$\pm$0.09)$\times10^{-1}$&$>$43.2$\pm$5.5&$<$0.18&--&$<$0.08$\pm$0.14\\
J162804.58-244838.0&(8.80$\pm$1.50$\pm$0.44)$\times10^{-2}$&$>$18.2$\pm$16.4&--&--&--\\
J162804.65-243456.6&(2.66$\pm$0.29$\pm$0.13)$\times10^{-1}$&$>$76.6$\pm$1.3&(1.60$\pm$0.24$\pm$0.08)$\times10^{-1}$&$>$58.5$\pm$3.8&-1.03$\pm$0.40\\
J162807.21-243040.9&(1.39$\pm$0.15$\pm$0.07)$\times10^{-1}$&41.2$\pm$26.9&--&--&--\\
J162807.28-244201.3&(8.50$\pm$1.50$\pm$0.43)$\times10^{-2}$&$>$45.0$\pm$12.8&--&--&--\\
J162813.72-244008.6&(8.49$\pm$0.83$\pm$0.42)$\times10^{-1}$&22.3$\pm$13.2&--&--&--\\
J162820.60-242546.1&(1.26$\pm$0.03$\pm$0.06)$\times10^{+1}$&35.2$\pm$2.3&--&--&--\\
J163012.16-243345.2&(2.57$\pm$0.06$\pm$0.13)$\times10^{+0}$&65.4$\pm$1.3&--&--&--\\
J163027.69-243300.2&(2.40$\pm$0.29$\pm$0.12)$\times10^{-1}$&55.3$\pm$8.2&(2.02$\pm$0.20$\pm$0.10)$\times10^{-1}$&45.5$\pm$11.8&-0.35$\pm$0.35\\
J163031.02-243158.2&(2.82$\pm$0.56$\pm$0.14)$\times10^{-1}$&5.0$\pm$17.0&(1.91$\pm$0.25$\pm$0.10)$\times10^{-1}$&$>$59.4$\pm$4.4&-0.79$\pm$0.50\\
J163032.26-243127.9&(7.88$\pm$0.22$\pm$0.39)$\times10^{-1}$&42.5$\pm$5.3&(9.22$\pm$0.60$\pm$0.46)$\times10^{-1}$&43.5$\pm$6.2&0.32$\pm$0.20\\
J163033.26-243038.7&(1.69$\pm$0.28$\pm$0.08)$\times10^{-1}$&$>$60.7$\pm$5.9&(2.20$\pm$0.75$\pm$0.11)$\times10^{-1}$&$>$24.1$\pm$12.1&0.53$\pm$0.78\\
J163033.64-243519.0&(1.32$\pm$0.19$\pm$0.07)$\times10^{-1}$&$>$10.6$\pm$24.5&(1.51$\pm$0.16$\pm$0.08)$\times10^{-1}$&38.1$\pm$19.6&0.27$\pm$0.39\\
J163035.21-243417.8&(1.94$\pm$0.27$\pm$0.10)$\times10^{-1}$&21.3$\pm$21.7&(1.33$\pm$0.15$\pm$0.07)$\times10^{-1}$&45.2$\pm$11.5&-0.76$\pm$0.39\\
J163035.63-243418.9&(1.03$\pm$0.04$\pm$0.05)$\times10^{+0}$&33.3$\pm$6.6&(1.07$\pm$0.03$\pm$0.05)$\times10^{+0}$&16.6$\pm$8.3&0.08$\pm$0.18\\
J163036.26-243135.3&(7.40$\pm$0.19$\pm$0.37)$\times10^{-1}$&41.4$\pm$4.7&(5.66$\pm$0.31$\pm$0.28)$\times10^{-1}$&36.7$\pm$9.1&-0.54$\pm$0.19\\
J163036.93-241334.9&(6.80$\pm$0.44$\pm$0.34)$\times10^{-1}$&17.0$\pm$9.7&(4.50$\pm$0.23$\pm$0.23)$\times10^{-1}$&32.3$\pm$9.5&-0.83$\pm$0.22\\
J163037.85-241206.0&(1.41$\pm$0.03$\pm$0.07)$\times10^{+0}$&15.1$\pm$6.2&(8.07$\pm$0.21$\pm$0.40)$\times10^{-1}$&18.1$\pm$8.4&-1.13$\pm$0.16\\
J163058.02-243441.3&(3.43$\pm$0.38$\pm$0.17)$\times10^{-1}$&29.4$\pm$12.7&(3.78$\pm$0.64$\pm$0.19)$\times10^{-1}$&$>$26.3$\pm$8.7&0.2$\pm$0.43\\
J163100.40-241640.4&(8.44$\pm$0.42$\pm$0.42)$\times10^{-1}$&45.7$\pm$4.5&--&--&--\\
J163109.63-242554.7&(1.07$\pm$0.03$\pm$0.05)$\times10^{+0}$&27.9$\pm$5.9&(9.64$\pm$0.75$\pm$0.48)$\times10^{-1}$&$>$42.0$\pm$6.2&-0.21$\pm$0.22\\
J163109.79-243008.4&(1.49$\pm$0.02$\pm$0.07)$\times10^{+0}$&21.1$\pm$4.2&(5.81$\pm$0.35$\pm$0.29)$\times10^{-1}$&27.1$\pm$7.2&-1.9$\pm$0.19\\
J163112.95-242722.6&(4.13$\pm$0.28$\pm$0.21)$\times10^{-1}$&17.2$\pm$12.2&(1.99$\pm$0.17$\pm$0.10)$\times10^{-1}$&$>$33.3$\pm$9.5&-1.48$\pm$0.26\\
J163115.01-243243.9&(8.20$\pm$0.46$\pm$0.41)$\times10^{-1}$&78.9$\pm$2.2&(6.31$\pm$0.36$\pm$0.32)$\times10^{-1}$&81.9$\pm$2.1&-0.53$\pm$0.22\\
J163115.25-243313.8&(2.10$\pm$0.21$\pm$0.10)$\times10^{-1}$&25.6$\pm$13.8&(1.75$\pm$0.19$\pm$0.09)$\times10^{-1}$&17.3$\pm$16.4&-0.37$\pm$0.33\\
J163115.75-243402.8&(4.57$\pm$0.30$\pm$0.23)$\times10^{-1}$&$>$79.2$\pm$0.7&(4.82$\pm$0.28$\pm$0.24)$\times10^{-1}$&$>$87.3$\pm$1.0&0.11$\pm$0.23\\
J163120.14-242928.5&(9.89$\pm$0.09$\pm$0.49)$\times10^{+0}$&11.8$\pm$3.5&(6.23$\pm$0.11$\pm$0.31)$\times10^{+0}$&11.5$\pm$3.0&-0.93$\pm$0.15\\
J163120.18-243001.0&(2.95$\pm$0.27$\pm$0.15)$\times10^{-1}$&59.3$\pm$6.5&(2.85$\pm$0.21$\pm$0.14)$\times10^{-1}$&57.2$\pm$8.3&-0.07$\pm$0.28\\
J163130.62-243351.6&(7.40$\pm$0.33$\pm$0.37)$\times10^{-1}$&8.3$\pm$7.6&(5.03$\pm$0.37$\pm$0.25)$\times10^{-1}$&50.0$\pm$6.7&-0.78$\pm$0.22\\
J163131.09-242719.6&(4.30$\pm$0.13$\pm$0.21)$\times10^{+0}$&44.2$\pm$3.0&(3.46$\pm$0.11$\pm$0.17)$\times10^{+0}$&33.9$\pm$3.2&-0.44$\pm$0.17\\
J163138.57-253220.0&(4.87$\pm$0.48$\pm$0.24)$\times10^{-1}$&61.3$\pm$6.4&--&--&--\\
J163140.49-245234.7&(7.68$\pm$0.37$\pm$0.38)$\times10^{-1}$&38.9$\pm$6.9&(4.67$\pm$0.42$\pm$0.23)$\times10^{-1}$&--\tablenotemark{a}&-1.01$\pm$0.25\\
J163140.67-241516.4&(9.20$\pm$0.62$\pm$0.46)$\times10^{-1}$&71.2$\pm$2.9&$<$0.47&--&$<$-1.36$\pm$0.14\\
J163151.93-245617.4&(2.87$\pm$0.13$\pm$0.14)$\times10^{+0}$&93.3$\pm$0.7&(4.43$\pm$0.31$\pm$0.22)$\times10^{+0}$&97.5$\pm$0.4&0.88$\pm$0.22\\
J163152.10-245615.7&(3.63$\pm$0.37$\pm$0.18)$\times10^{-1}$&$>$67.2$\pm$5.4&(1.26$\pm$0.09$\pm$0.06)$\times10^{+0}$&$>$41.1$\pm$4.8&2.52$\pm$0.29\\
J163152.34-253144.7&(1.50$\pm$0.19$\pm$0.07)$\times10^{-1}$&16.5$\pm$19.1&(1.12$\pm$0.22$\pm$0.06)$\times10^{-1}$&$>$34.8$\pm$8.8&-0.59$\pm$0.49\\
J163154.49-245217.1&(6.46$\pm$0.17$\pm$0.32)$\times10^{-1}$&35.5$\pm$4.7&(3.53$\pm$0.27$\pm$0.18)$\times10^{-1}$&30.8$\pm$15.4&-1.22$\pm$0.22\\
J163159.36-245639.7&(2.19$\pm$0.04$\pm$0.11)$\times10^{+0}$&10.8$\pm$5.2&(1.56$\pm$0.07$\pm$0.08)$\times10^{+0}$&26.1$\pm$6.2&-0.69$\pm$0.17\\
J163159.51-252918.7&(3.28$\pm$0.21$\pm$0.16)$\times10^{-1}$&23.7$\pm$11.9&(2.52$\pm$0.17$\pm$0.13)$\times10^{-1}$&5.6$\pm$17.5&-0.53$\pm$0.24\\
J163200.97-245643.3&(7.19$\pm$0.37$\pm$0.36)$\times10^{-1}$&52.7$\pm$5.8&(7.94$\pm$0.27$\pm$0.40)$\times10^{-1}$&43.7$\pm$3.6&0.2$\pm$0.19\\
J163202.39-245710.0&(2.86$\pm$0.40$\pm$0.14)$\times10^{-1}$&$>$67.4$\pm$3.6&(2.15$\pm$0.39$\pm$0.11)$\times10^{-1}$&$>$42.5$\pm$5.3&-0.58$\pm$0.48\\
J163204.79-245636.8&(4.35$\pm$0.53$\pm$0.22)$\times10^{-1}$&34.3$\pm$12.4&$<$0.12&--&$<$-2.6$\pm$0.14\\
J163208.05-253016.3&(6.08$\pm$0.08$\pm$0.30)$\times10^{+0}$&2.2$\pm$5.9&(4.11$\pm$0.09$\pm$0.21)$\times10^{+0}$&5.9$\pm$8.1&-0.79$\pm$0.15\\
J163210.20-245618.9&(3.12$\pm$0.41$\pm$0.16)$\times10^{-1}$&$>$44.3$\pm$5.5&(2.04$\pm$0.13$\pm$0.10)$\times10^{-1}$&--\tablenotemark{a}&-0.86$\pm$0.33\\
J163210.77-243827.6&(5.33$\pm$0.33$\pm$0.27)$\times10^{-1}$&23.1$\pm$7.7&(3.61$\pm$0.27$\pm$0.18)$\times10^{-1}$&57.6$\pm$9.2&-0.79$\pm$0.24\\
J163210.96-253021.3&(1.23$\pm$0.16$\pm$0.06)$\times10^{+0}$&--&$<$0.13&--&Extended\\
J163211.08-243651.1&(7.69$\pm$0.27$\pm$0.38)$\times10^{-1}$&38.7$\pm$5.9&(6.62$\pm$0.82$\pm$0.33)$\times10^{-1}$&$>$25.5$\pm$14.9&-0.3$\pm$0.30\\
J163211.79-244021.8&(1.50$\pm$0.03$\pm$0.08)$\times10^{+0}$&28.0$\pm$3.4&(1.32$\pm$0.05$\pm$0.07)$\times10^{+0}$&30.0$\pm$6.8&-0.27$\pm$0.16\\
J163212.25-243643.7&(6.03$\pm$0.22$\pm$0.30)$\times10^{-1}$&28.7$\pm$8.0&$<$0.26&--&$<$-1.7$\pm$0.14\\
J163213.92-244407.8&(3.37$\pm$0.53$\pm$0.17)$\times10^{-1}$&$>$54.5$\pm$7.6&--&--&--\\
J163214.16-252344.5&(2.44$\pm$0.06$\pm$0.12)$\times10^{+1}$&29.2$\pm$1.7&--&--&--\\
J163227.41-243951.4&(6.82$\pm$0.25$\pm$0.34)$\times10^{-1}$&20.4$\pm$5.1&(4.45$\pm$0.21$\pm$0.22)$\times10^{-1}$&37.2$\pm$5.8&-0.86$\pm$0.19\\
J163231.17-244014.6&(5.10$\pm$0.27$\pm$0.26)$\times10^{-1}$&34.3$\pm$5.9&(4.16$\pm$0.25$\pm$0.21)$\times10^{-1}$&32.5$\pm$6.6&-0.41$\pm$0.22\\
J163245.23-243647.4&(7.08$\pm$0.57$\pm$0.35)$\times10^{-1}$&51.0$\pm$7.8&--&--&--\\
J163421.10-235625.1&(1.80$\pm$0.20$\pm$0.09)$\times10^{-1}$&$>$76.1$\pm$1.2&(1.48$\pm$0.16$\pm$0.07)$\times10^{-1}$&$>$73.7$\pm$2.5&-0.4$\pm$0.34\\
J163436.01-235614.5&(8.57$\pm$0.56$\pm$0.43)$\times10^{-1}$&19.7$\pm$9.5&(4.72$\pm$0.77$\pm$0.24)$\times10^{-1}$&33.2$\pm$12.5&-1.21$\pm$0.38\\
J163437.30-235946.2&(5.88$\pm$0.34$\pm$0.29)$\times10^{-1}$&17.5$\pm$8.2&(7.37$\pm$0.56$\pm$0.37)$\times10^{-1}$&$>$33.1$\pm$8.4&0.46$\pm$0.24\\
J163551.89-242253.6&(5.44$\pm$0.50$\pm$0.27)$\times10^{-1}$&45.5$\pm$7.5&--&--&--\\
J163557.74-241447.9&(1.77$\pm$0.06$\pm$0.09)$\times10^{+0}$&32.8$\pm$8.5&--&--&--\\
J163615.79-242159.8&(2.61$\pm$0.36$\pm$0.13)$\times10^{-1}$&15.3$\pm$10.6&(2.12$\pm$0.14$\pm$0.11)$\times10^{-1}$&7.9$\pm$14.2&-0.42$\pm$0.34\\
J163617.50-242555.4&(2.06$\pm$0.07$\pm$0.10)$\times10^{+0}$&11.4$\pm$7.0&(1.47$\pm$0.17$\pm$0.07)$\times10^{+0}$&$>$29.7$\pm$7.4&-0.68$\pm$0.28\\
J163626.93-242117.8&(1.76$\pm$0.20$\pm$0.09)$\times10^{-1}$&29.8$\pm$15.7&(2.00$\pm$0.31$\pm$0.10)$\times10^{-1}$&$>$35.9$\pm$10.4&0.26$\pm$0.41\\
J163639.40-241710.3&(1.26$\pm$0.06$\pm$0.06)$\times10^{+0}$&46.0$\pm$6.1&--&--&--\\
J163949.54-235939.0&(2.62$\pm$0.41$\pm$0.13)$\times10^{-1}$&32.9$\pm$19.9&(1.90$\pm$0.27$\pm$0.10)$\times10^{-1}$&$>$25.2$\pm$8.0&-0.65$\pm$0.45\\
J164002.06-240137.0&(3.96$\pm$0.26$\pm$0.20)$\times10^{-1}$&19.2$\pm$10.5&(2.15$\pm$0.24$\pm$0.11)$\times10^{-1}$&$>$10.3$\pm$14.1&-1.23$\pm$0.30\\

\enddata
\tablenotetext{a}{Source not detected at three times the noise level on individual epochs,
but detected on the image of the concatenated epochs.}
\end{deluxetable}

\begin{deluxetable}{ccrcrc}
\tabletypesize{\scriptsize}
\tablewidth{0pt}
\tablecolumns{6}
\tablecaption{Sources detected in circular polarization}
\tablehead{{GB-VLA Name} & Source type & X polz (\%) & C polz (\%)\\}
\startdata
J162557.51-243032.1  & YSO/III                 & 22.1 (L)    &    $<$16.3\\
J162603.01-242336.4  & YSO/III                 & 6.5 (L)      &        4.7(L)\\
J162634.17-242328.4  & YSO/III                  & 5.8 (L)     &        --\\
J162752.08-244050.5  & YSO            & 11.2 (L)   &   8.9 (L)\\
J163115.01-243243.9  & YSO/WTT            &16.2 (L)    & 8.3(L)\\
J163200.97-245643.3  & YSO/I                    & 21.4 (R)  &  16.1(R)\\
J163211.79-244021.8  & YSO/WTT/II         & --               & 16.5(L)\\
\enddata
\label{tab:polz}
\end{deluxetable}

\begin{deluxetable}{lccccccc}
\tabletypesize{\scriptsize}
\tablewidth{0pt}
\tablecolumns{8}
\tablecaption{Radio Sources with known counterparts \label{tab:counterparts}}
\tablehead{           & Other         &         & \multicolumn{3}{c}{Infrared\tablenotemark{b,c}} &       & Object\\
\colhead{GB-VLA Name} & \colhead{Names} &\colhead{X-ray\tablenotemark{a}} & \colhead{SST} & \colhead{2M}&\colhead{WISE}&
\colhead{Radio\tablenotemark{d}}&\colhead{type}\\}
\startdata
J162540.94-244147.2&SSTc2d J162540.9-244147&--
&Y&--&--&--
&--
\\
J162542.48-242143.7&--&--
&--&--&--&ROC 4
&--
\\
J162548.96-244049.5&SSTc2d J162549.0-244049&--
&Y&--&--&ROC 6
&--
\\
J162550.51-243914.7&WLY 2-10 
&ROXRA 5
&Y&Y&Y&--
&YSO
\\
J162556.09-243015.3&WLY 2-11
&--
&Y&Y&Y&GDS J162556.1-243014
&YSO
\\
J162557.51-243032.1&YLW 24 
&ROXRA 7
&Y&Y&Y&--
&YSO
\\
J162603.01-242336.4&DoAr 21
&A-2
&Y&Y&Y&ROC 8
&YSO
\\
J162607.63-242741.7&BKLT J162607-242742
&A-5
&Y&Y&Y&--
&YSO
\\
J162608.04-242523.1&--&--
&--&--&--&ROC 9
&--
\\
J162610.32-242054.9&GSS 26
&A-6
&Y&Y&Y&GDS J162610.3-242054&YSO
\\
J162610.55-242853.3&--&--
&--&--&--&GDS J162610.5-242853
&--
\\
J162611.08-242907.4&GDS J162611.0-242908 &--
&--&--&--&GDS J162611.0-242908
&--
\\
J162615.67-243421.2&FG Oph 18
&--
&--&--&--&NVSS 162615-243419
&E
\\
J162616.84-242223.5&BKLT J162616-242225
&A-14
&Y&Y&Y&LFAM p1 
&YSO
\\
J162621.72-242250.7&SSTc2d J162621.7-242250
&--
&Y&--&--&LFAM 1
&YSO
\\
J162622.38-242253.3&BKLT J162622-242254 
&A-20
&Y&Y&--&LFAM 2
&YSO
\\
J162623.57-242439.6&BKLT J162623-242441 
&A-23
&Y&Y&Y&LFAM 3
&YSO
\\
J162624.08-241613.5&YLW 32
&A-25
&Y&Y&Y&GDS J162624.0-241613 
&YSO
\\
J162625.62-242429.2&VLA 1623W
&--
&Y&--&--&LFAM 4
&YSO
\\
J162626.03-244923.7&SFAM 11
&--
&Y&--&--&ROC12
&E
\\
J162626.31-242430.2&VLA 1623B
&--
&--&--&--&(?) Knot of VLA 1623
&YSO?
\\
J162626.39-242430.9&VLA 1623A
&--
&Y&--&--&VLA 1623
&YSO
\\
J162629.63-242317.2&--&--
&--&--&--&GDS J162629.5-242317
&--
\\
J162629.67-241905.8&BKLT J162629-241908
&A-33
&Y&Y&Y&LFAM 8
&YSO
\\
J162630.59-242023.0&GDS J162630.6-242023
&A-35
&--&--&--&--
&--
\\
J162631.28-241832.9&GDS J162631.3-241833
&A-37
&--&--&--&LFAM 10
&--
\\
J162633.48-241215.9&GDS J162633.4-241216
&--
&--&--&--&SFAM 12
&--
\\
J162634.17-242328.4&S1
&A-41
&Y&Y&--&SFAM 13
&YSO
\\
J162634.95-242655.3&--&--
&--&--&--&LFAM 12
&--
\\
J162635.33-242405.2&GDS J162635.3-242405
&--
&--&--&--&LFAM 13
&--
\\
J162639.00-243052.8&--&--
&--&--&--&GDS J162639.0-243052
&--
\\
J162642.44-242626.1&BKLT J162642-242627
&A-51
&Y&Y&Y&LFAM 15
&YSO
\\
J162642.53-244628.3&SSTc2d J162642.5-244628&--
&Y&--&--&--
&--
\\
J162643.76-241633.4&VSSG 11
&A-56
&Y&Y&Y&SFAM 15
&YSO
\\
J162646.36-242002.0&GDS J162646.3-242001
&--
&--&--&--&LFAM 17
&--
\\
J162647.17-245157.4&SSTc2d J162647.2-245157&--
&Y&--&--&--
&--
\\
J162649.23-242003.3&VSSG 3
&A-63
&Y&Y&Y&LFAM 18 
&YSO
\\
J162651.69-241441.5&SSTc2d J162651.7-241441&--
&Y&Y&Y&--
&--
\\
J162653.38-241105.8&SSTc2d J162653.4-241106&--
&Y&--&--&--
&--
\\
J162658.25-243738.5&BKLT J162658-243739A
&--
&--&--&--&--
&--
\\
J162658.38-242130.5&YLW 1C
&A-72
&Y&Y&Y&--
&YSO
\\
J162659.16-243458.9&YLW 4C
&DROXO 27
&Y&--&--&LFAM 23
&YSO
\\
J162700.00-242640.3&FG Oph 21
&A-75
&Y&--&--&LFAM 21
&E
\\
J162700.02-243537.6&--&--
&--&--&--&LFAM 24
&--
\\
J162702.11-243842.5&--&--
&--&--&--&LFAM 25
&--
\\
J162702.15-241927.8&GDS J162702.1-241928
&--
&--&--&--&GDS J162702.1-241928 
&--
\\
J162702.33-243727.3&YLW 5
&BF-12
&Y&--&--&--
&YSO
\\
J162702.36-242724.8&SSTc2d J162702.4-242725&--
&Y&--&--&GDS J162702.3-242724
&--
\\
J162705.16-242007.8&VSSG 21
&A-80
&Y&Y&Y&--
&YSO
\\
J162705.25-243629.8&ISO-Oph 99
&--
&Y&Y&Y&LFAM 26
&YSO
\\
J162705.96-242618.9&YLW 10A
&A-81
&Y&Y&Y&--
&YSO
\\
J162709.41-243719.0&YLW 7
&DROXO 38
&Y&Y&Y&LFAM 27 
&YSO
\\
J162711.29-243722.5&--&--
&--&--&--&LFAM 28
&--
\\
J162713.62-242226.1&SSTc2d J162713.6-242226&--
&Y&--&--&LFAM 32
&--
\\
J162714.71-243919.7&--&--
&--&--&--&LFAM 29
&--
\\
J162715.69-243845.7&WL 20S
&--
&Y&Y&Y&--
&YSO
\\
J162717.38-243616.7&--&--
&--&--&--&LFAM 32
&--
\\
J162718.17-242852.9&YLW 12B
&DROXO 49
&Y&Y&--&SFAM 22
&YSO
\\
J162719.34-243130.4&--&BF-45
&--&--&--&--
&--
\\
J162719.50-244140.3&V2247 Oph
&DROXO 53
&Y&Y&Y&LFAM p6
&YSO
\\
J162721.81-244335.9&[GY92] 253
&DROXO 55
&Y&Y&Y&--
&YSO
\\
J162721.97-242940.0&[GY92] 256
&BF-51
&Y&Y&--&--
&YSO
\\
J162722.96-242236.6&[MPK2010b] 3809
&--
&Y&Y&--&--&YSO
\\
J162724.19-242929.8&BKLT J162724-242929
&BF-55
&Y&Y&--&--
&YSO
\\
J162726.90-244050.8&YLW 15
&DROXO 62
&Y&Y&Y&LFAM 33
&YSO
\\
J162727.36-243116.8&BKLT J162727-243116
&DROXO 63
&Y&Y&Y&--
&YSO
\\
J162728.00-243933.7&YLW 16A
&DROXO 64
&Y&Y&Y&LFAM 35
&YSO
\\
J162729.23-241755.3&--&--
&--&--&--&ROC 25
&--
\\
J162730.82-244727.2&BKLT J162730-244726
&DROXO 71
&Y&Y&Y&--
&YSO
\\
J162731.05-243403.4&BKLT J162731-243402
&DROXO 72
&Y&Y&Y&--
&YSO
\\
J162731.19-242833.8&OphB2 S7
&--
&Y&--&--&--
&--
\\
J162732.68-243324.5&BKLT J162732-243323
&DROXO 75
&--&Y&Y&LFAM p7
&YSO
\\
J162734.55-242020.7&--&--
&--&--&--&ROC 26
&--
\\
J162735.13-242624.0&--&--
&--&--&--&OphB2 S8
&--
\\
J162736.00-241402.9&--&--
&--&--&--&ROC 27
&--
\\
J162738.20-242630.5&--&--
&--&--&--&OphB2 S9
&--
\\
J162739.42-243915.8&BKLT J162739-243914
&DROXO 86
&Y&Y&Y&LFAM p8
&YSO
\\
J162740.92-241452.8&--&--
&--&--&--&ROC 28
&--
\\
J162741.39-241454.5&--&--
&--&--&--&ROC 28
&--
\\
J162741.49-243537.6&BKLT J162741-243537
&DROXO 89
&Y&--&--&--
&YSO
\\
J162749.85-242540.5&YLW 53
&--
&Y&Y&Y&ROC 31
&YSO
\\
J162751.80-243145.9&YLW 52&DROXO 97 & Y&Y & Y&-- &YSO\\ 
J162751.89-244630.1&BKLT J162752-244630
&DROXO 98
&Y&Y&Y&--
&YSO
\\
J162752.08-244050.5&HBC 642
&DROXO 99
&Y&Y&Y&--
&YSO
\\
J162752.30-242929.9&SSTc2d J162752.3-242930&--
&Y&--&--&--
&--
\\
J162753.08-242830.3&SSTc2d J162753.1-242830&--
&Y&--&Y&--
&--
\\
J162757.81-244001.9&BKLT J162757-244004
&DROXO 101
&Y&Y&Y&--
&YSO
\\
J162759.09-242225.9&SSTc2d J162759.1-242226&--
&Y&--&--&--
&--
\\
J162759.95-244819.5&BKLT J162800-244819
&DROXO 102
&Y&Y&Y&--
&YSO
\\
J162803.39-242124.8&ROC 33
&--
&Y&--&--&ROS 15 
&--
\\
J162803.51-242131.2&--&--
&--&--&--&SFAM 30
&--
\\
J162804.65-243456.6&BKLT J162804-243459
&DROXO 105
&Y&Y&Y&--
&YSO
\\
J162807.21-243040.9&SSTc2d J162807.2-243041&--
&Y&--&--&--
&--
\\
J162813.72-244008.6&SFAM 34
&--
&--&--&--&ROC 35&--
\\
J162820.60-242546.1&SSTc2d J162820.6-242546&--
&Y&--&Y&ROC 37
&--
\\
J163012.16-243345.2&SSTc2d J163012.2-243345&--
&Y&--&--&ROC 43
&--
\\
J163027.69-243300.2&SSTc2d J163027.7-243300&--
&Y&--&--&--
&--
\\
J163032.26-243127.9&SSTc2d J163032.3-243128&--
&Y&--&--&--
&--
\\
J163033.26-243038.7&SSTc2d J163033.2-243039&--
&Y&--&--&--
&--
\\
J163033.64-243519.0&SSTc2d J163033.6-243519&--
&Y&--&--&--
&--
\\
J163035.63-243418.9&2E 3707 
&ROX 39
&Y&Y&Y&SFAM 87
&YSO
\\
J163036.93-241334.9&SSTc2d J163036.9-241335&--
&Y&--&Y&--
&--
\\
J163037.85-241206.0&SFAM 90
&--
&Y&--&Y&ROS 36
&--
\\
J163100.40-241640.4&SSTc2d J163100.4-241640&--
&Y&--&--&SFAM 105
&--
\\
J163109.63-242554.7&SSTc2d J163109.6-242555&--
&Y&--&--&--
&--
\\
J163109.79-243008.4&ROC 49
&--
&--&--&--&SFAM 108
&--
\\
J163115.01-243243.9&2MASS J16311501-2432436
&--
&Y&Y&Y&--
&YSO
\\
J163115.75-243402.8&IRAS 16282-2427 
&--
&Y&Y&Y&--
&YSO
\\
J163120.14-242928.5&ROC 52
&--
&--&--&--&SFAM 112
&--
\\
J163120.18-243001.0&NTTS 162819-2423N
&ROXs 43B
&Y&Y&--&--
&YSO
\\
J163130.62-243351.6&SSTc2d J163130.6-243352&--
&Y&--&--&--
&--
\\
J163131.09-242719.6&6dFGS gJ163131.1-242720
&--
&Y&Y&Y&ROC 53
&E
\\
J163140.49-245234.7&SSTc2d J163140.5-245235&--
&Y&--&--&SFAM 122
&--
\\
J163152.10-245615.7&LDN 1689 IRS 5
&--
&Y&Y&Y&L1689S1 1 
&YSO
\\
J163154.49-245217.1&SSTc2d J163154.5-245217&--
&Y&--&--&--
&--
\\
J163159.36-245639.7&L1689S1 2
&--
&--&--&--&SFAM 127
&--
\\
J163159.51-252918.7&SSTc2d J163159.5-252919&--
&Y&--&--&--
&--
\\
J163200.97-245643.3&WLY 2-67
&--
&Y&Y&Y&L1689S1 3 
&YSO
\\
J163208.05-253016.3&--&--
&--&--&--&SFAM 130
&--
\\
J163210.77-243827.6&SSTc2d J163210.8-243827&--
&Y&--&Y&--
&--
\\
J163211.08-243651.1&SSTc2d J163211.1-243652&--
&Y&--&--&--
&--
\\
J163211.79-244021.8&V2248 Oph
&1RXS J163212.8-244013
&Y&Y&Y&--
&YSO
\\
J163214.16-252344.5&SSTc2d J163214.1-252345&--
&Y&--&--&NVSS 163214-252344
&--
\\
J163227.41-243951.4&SSTc2d J163227.4-243951&--
&Y&--&--&--
&--
\\
J163231.17-244014.6&SSTc2d J163231.2-244014&--
&Y&--&--&--
&--
\\
J163245.23-243647.4&SFAM 200
&--
&Y&--&--&--
&--
\\
J163421.10-235625.1&WSB 80 
&RX J1634.3-2356
&--&Y&Y&--
&YSO
\\
J163436.01-235614.5&--&--
&--&--&--&NVSS 163436-235611
&--
\\
J163551.89-242253.6&SSTc2d J163551.9-242253&--
&Y&--&--&--
&--
\\
J163557.74-241447.9&--&--
&--&--&--&NVSS 163557-241446
&--
\\
J163615.79-242159.8&SSTc2d J163615.8-242160&--
&Y&--&--&--
&--
\\
J163617.50-242555.4&SSTc2d J163617.5-242555&--
&Y&--&--&SFAM 212
&--
\\
J163639.40-241710.3&SSTc2d J163639.4-241710&--
&Y&--&Y&--
& --
\\
J163949.54-235939.0&SSTc2d J163949.5-235939&--
&Y&--&Y&--
&--
\\

\enddata
\tablenotetext{a}{ROXRA = Grosso et al.\ (2000); A and BF = Imanishi et al.\ (2003);
DROXO = Pillitteri et al.\ (2010); ROX = Montmerle et al.\ (1983); ROXs = Bouvier \&
Appenzeller\ (1992); 1RXS = Voges et al.\ (1999) and RX = Martin et al.\ (1998).
}\tablenotetext{b}{SST = Evans et al.\ (2009); 2M = Cutri et al.\ (2003) and WISE = Cutri
et al\ (2012).}
\tablenotetext{c}{For GBDS VLA J162722.96-242236.6, 2MASS and SST data from Marsh et al.\ (2010).}
\tablenotetext{d}{ROC and ROS = Andr\'e et al.\ (1987); GDS = Gagne et al.\ (2003);
NVSS = Condon et al.\ (1998); LFAM = Leous et al.\ (1991); SFAM = Stine et al.\ (1988)}
\end{deluxetable}

\begin{deluxetable}{ccrcrc}
\tabletypesize{\scriptsize}
\tablewidth{0pt}
\tablecolumns{4}
\tablecaption{Young stellar object candidates based just in their radio properties \label{tab:possibleYSOs}}
\tablehead{           & \multicolumn{2}{c}{Flux Properties}	&Spectral\\
\colhead{GB-VLA Name} &\colhead{Var.$_{\rm 4.5}$\,(\%)} & \colhead{Var.$_{\rm 7.5}$\,(\%)}&\colhead{Index}\\
}\startdata
J162605.29-243436.6&--&33.9$\pm$27.2&$>$1.34$\pm$0.14\\
J162632.78-241627.5&38.8$\pm$18.7&32.1$\pm$18.4&0.86$\pm$0.49\\
J162633.16-245246.7&30.7$\pm$5.5&49.3$\pm$5.3&0.27$\pm$0.18\\
J162637.27-244553.9&52.8$\pm$8.4&$>$78.0$\pm$3.0&-0.96$\pm$0.40\\
J162641.11-245855.8&67.8$\pm$3.3&81.0$\pm$2.3&0.5$\pm$0.18\\
J162646.36-242002.0&15.3$\pm$6.4&21.6$\pm$5.7&0.66$\pm$0.17\\
J162647.23-243620.3&21.5$\pm$6.9&22.0$\pm$7.9&0.31$\pm$0.19\\
J162702.11-243842.5&$>$55.5$\pm$4.3&68.6$\pm$7.3&0.04$\pm$0.26\\
J162752.30-242929.9&$>$43.8$\pm$14.8&$>$17.1$\pm$16.9&0.38$\pm$0.48\\
J162756.01-244810.7&--&$>$55.8$\pm$15.9&$>$1.53$\pm$0.14\\
J162803.51-242131.2&$>$43.2$\pm$5.5&--&$<$0.08$\pm$0.14\\
J163032.26-243127.9&42.5$\pm$5.3&43.5$\pm$6.2&0.32$\pm$0.20\\
J163033.26-243038.7&$>$60.7$\pm$5.9&$>$24.1$\pm$12.1&0.53$\pm$0.78\\
J163033.64-243519.0&$>$10.6$\pm$24.5&38.1$\pm$19.6&0.27$\pm$0.39\\
J163058.02-243441.3&29.4$\pm$12.7&$>$26.3$\pm$8.7&0.2$\pm$0.43\\
J163151.93-245617.4&93.3$\pm$0.7&97.5$\pm$0.4&0.88$\pm$0.22\\
J163437.30-235946.2&17.5$\pm$8.2&$>$33.1$\pm$8.4&0.46$\pm$0.24\\
J163626.93-242117.8&29.8$\pm$15.7&$>$35.9$\pm$10.4&0.26$\pm$0.41\\

\enddata
\end{deluxetable}

\begin{deluxetable}{lccccccc}
\tabletypesize{\scriptsize}
\tablewidth{0pt}
\tablecolumns{8}
\tablecaption{Young Stellar Objects detected in the radio observations \label{tab:yso}}
\tablehead{           & Spectral       &  SED                    &  &  & & & \\
\colhead{GB-VLA Name} & \colhead{type}& \colhead{clasification}& 
\colhead{Var.\tablenotemark{a}} & \colhead{Pol.\tablenotemark{a}} & \colhead{$\alpha$\tablenotemark{a}}&\colhead{X-ray} &\colhead{Ref.\tablenotemark{b}}\\}
\startdata
J162550.51-243914.7&K5.5&Class III&Y&N&N&Y&1\\
J162556.09-243015.3&M5&Class III&N&N&F&N&1\\
J162557.51-243032.1&K6&Class III&Y&Y&N&Y&1\\
J162603.01-242336.4&K0&Class III&Y&Y&F&Y&1,2\\
J162607.63-242741.7&--&Class III&Y&N&N&Y&1\\
J162610.32-242054.9&M0&Class II&Y&--&P&Y&1\\
J162616.84-242223.5&K6&Class II&Y&N&F&Y&1\\
J162621.72-242250.7&--&Class I&N&--&P&N&1\\
J162622.38-242253.3&K8&Class II&Y&N&N&Y&1\\
J162623.57-242439.6&M0&FS&N&--&F&Y&1\\
J162624.08-241613.5&K5.5&Class II&N&--&P&Y&1,2\\
J162625.62-242429.2&--&Class I&N&--&F&N&3\\
J162626.31-242430.3\tablenotemark{c}&--&Class 0 (?)&Y&--&P&N&3\\
J162626.39-242430.9&--&Class 0&N&--&P&N&4\\
J162629.67-241905.8&M1.5&Class III&Y&N&N&Y&1\\
J162634.17-242328.4&B4&Class III&N&Y&N&Y&1\\
J162642.44-242626.1&--&Class III&Y&N&F&Y&1\\
J162643.76-241633.4&M0&Class III&N&--&F&Y&1,2\\
J162649.23-242003.3&K6&Class III&Y&--&N&Y&1\\
J162658.38-242130.5&--&Class II&Y&--&N&Y&1\\
J162659.16-243458.9&--&Class I&N&N&P&Y&1\\
J162702.33-243727.3&B8-A7&Class I&N&--&P&Y&1\\
J162705.16-242007.8&--&Class III&N&--&N&Y&1\\
J162705.25-243629.8&--&Class I&N&--&F&N&1\\
J162705.96-242618.9&--&Class II&Y&--&N&Y&1\\
J162709.41-243719.0&--&Class I&N&N&P&Y&1\\
J162715.69-243845.7&BD?&--&N&N&F&N&5\\
J162718.17-242852.9&F7&Class III&Y&--&N&Y&1\\
J162719.50-244140.4&M2.5&Class III&N&--&N&Y&1\\
J162721.81-244336.0&M3&Class III&Y&N&F&Y&1\\
J162721.97-242940.0&--&Class II&N&--&N&Y&1\\
J162722.96-242236.6&BD&--&Y&--&N&N&6\\
J162724.19-242929.6&--&Class III&Y&--&N&Y&1\\
J162726.90-244050.8&--&Class I&N&--&N&Y&1\\
J162727.36-243116.8&M0&Class II&Y&N&N&Y&1\\
J162728.00-243933.7&K8&Class I&N&N&F&Y&1\\
J162730.82-244727.2&M1&Class III&Y&--&N&Y&5,7\\
J162731.05-243403.4&--&Class III&Y&N&P&Y&1\\
J162732.68-243324.5&M2&Class II&Y&N&F&Y&8\\
J162739.41-243915.8&K5&Class II&N&--&P&Y&1\\
J162741.49-243537.6&--&Class III&Y&--&N&Y&1\\
J162749.85-242540.5&A7&Class III&N&Y&F&N&1\\
J162751.80-243145.9&--&Class I &N &N &N &Y &7 \\
J162751.89-244630.1&--&Class III&Y&N&N&Y&7\\
J162752.08-244050.5&K7&--&Y&Y&P&Y&9\\
J162757.81-244001.9&K7&Class III&Y&--&N&Y&1,10\\
J162759.95-244819.5&M4.75&WTTS&Y&--&P&Y&1,11\\
J162804.65-243456.6&--&Class II&Y&--&N&Y&8\\
J163035.63-243418.9&K5&CTTS&N&N&F&Y&9,12\\
J163115.01-243243.9&M0&WTTS&Y&Y&N&N&9,13\\
J163115.75-243402.8&K6&WTTS&Y&N&F&N&9,12\\
J163120.18-243001.0&K3e&WTTS&Y&N&F&Y&9,14\\
J163152.10-245615.7&M3&FS&Y&--&P&N&8\\
J163200.97-245643.3&K5&Class I&Y&Y&P&N&8\\
J163211.79-244021.8&M3&WTTS/Class II&N&Y&N&Y&8\\
J163421.10-235625.1&M4.5&WTTS&Y&--&N&Y&11,9\\

\enddata
\tablenotetext{a}{Var.\ = Y when the source variability is higher than 50\% in at least one frequency; N when it is lower.Pol.\ = Y when circular polarization is detected and N when it is not. Sources in the outer fields for which the polarization could not be assessed are shown as "--". $\alpha$ refers to the spectral index, and is given as P (for positive) when it ishigher than 0.2; F (for flat) when it is between --0.2 and $+$0.2, and N (for negative) when is is lower than --0.2. X-ray\ = Y when there is a X-ray flux reported in literature, N when it is not.}
\tablenotetext{b}{1= Wilking et al.\ (2005); 2 = Gagne et al.\ (2004); 3 = Murillo \& Lai (2013);
4 = Andre et al.\ (1993); 5 = Alves de Oliveira et al.\ (2010); 6 = Marsh et al.\ (2010);
7 = Pilliteri et al.\ (2010); 8 = McClure et al.\ (2010); 9 = Cieza et al.\ (2007);
10 = Wilking et al.\ (2001); 11 = Mart\'i et al.\ (1998); 12 = Wahhaj et al.\ (2010);
13 = Bouvier \& Apenzeller\ (1992) and 14 = Torres et al.\ (2006)}
\tablenotetext{c}{This source corresponds to VLA1623B; as discussed in the text, it may be a young star or an outflow knot feature.}\end{deluxetable}

\begin{figure*}[b!]
\begin{center}
\includegraphics[width=0.80\textwidth,angle=0]{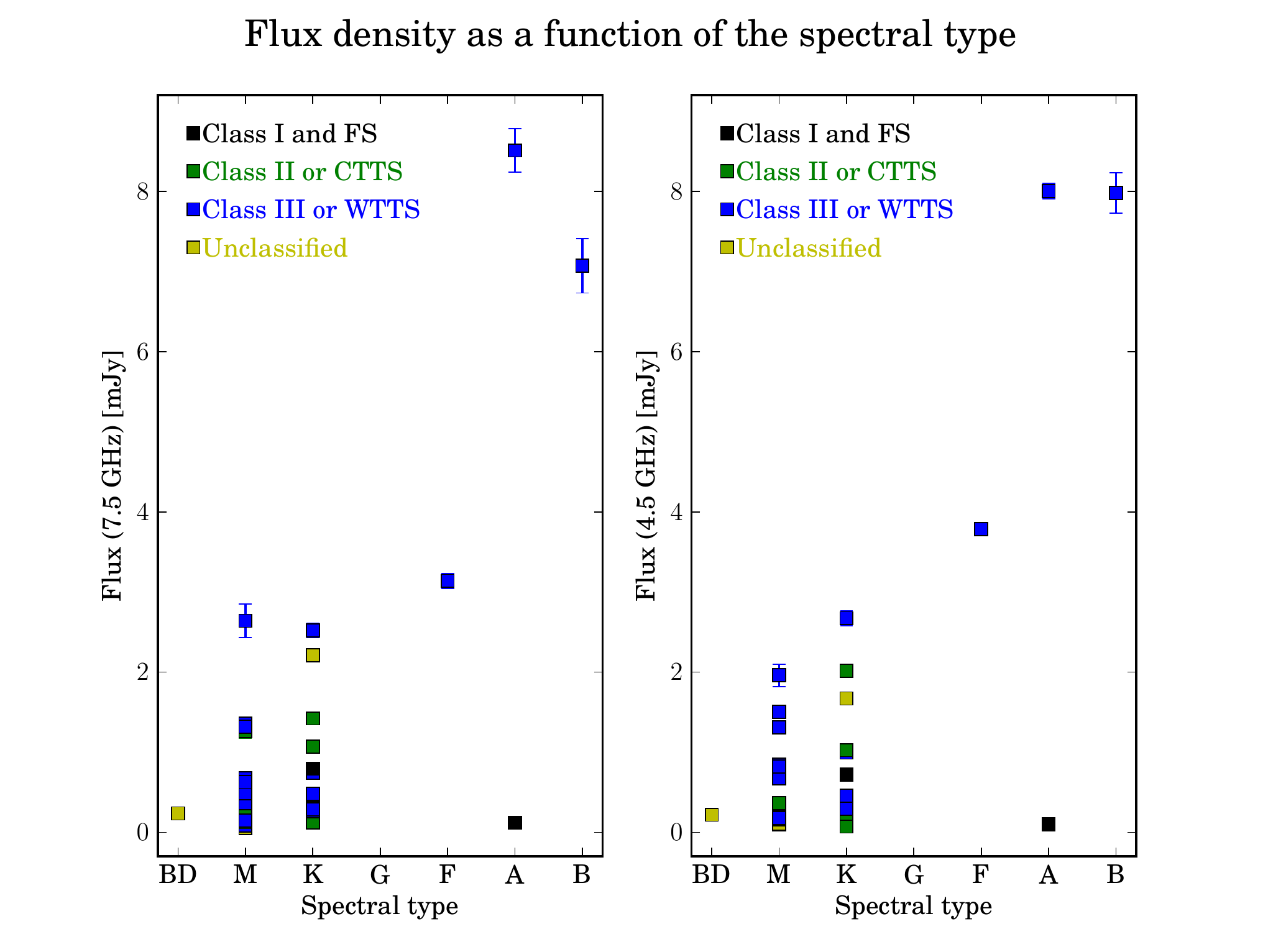}
\end{center}
\caption{Radio flux at 7.5 GHz (left) and 4.5 GHz (right) as a function of YSO 
spectral type. Colors indicate the evolutionary of the 
object as explained at the top-left of the diagrams.}
\label{fig:imf}
\end{figure*}

\begin{figure*}[b!]
\begin{center}
\includegraphics[width=0.80\textwidth,angle=0]{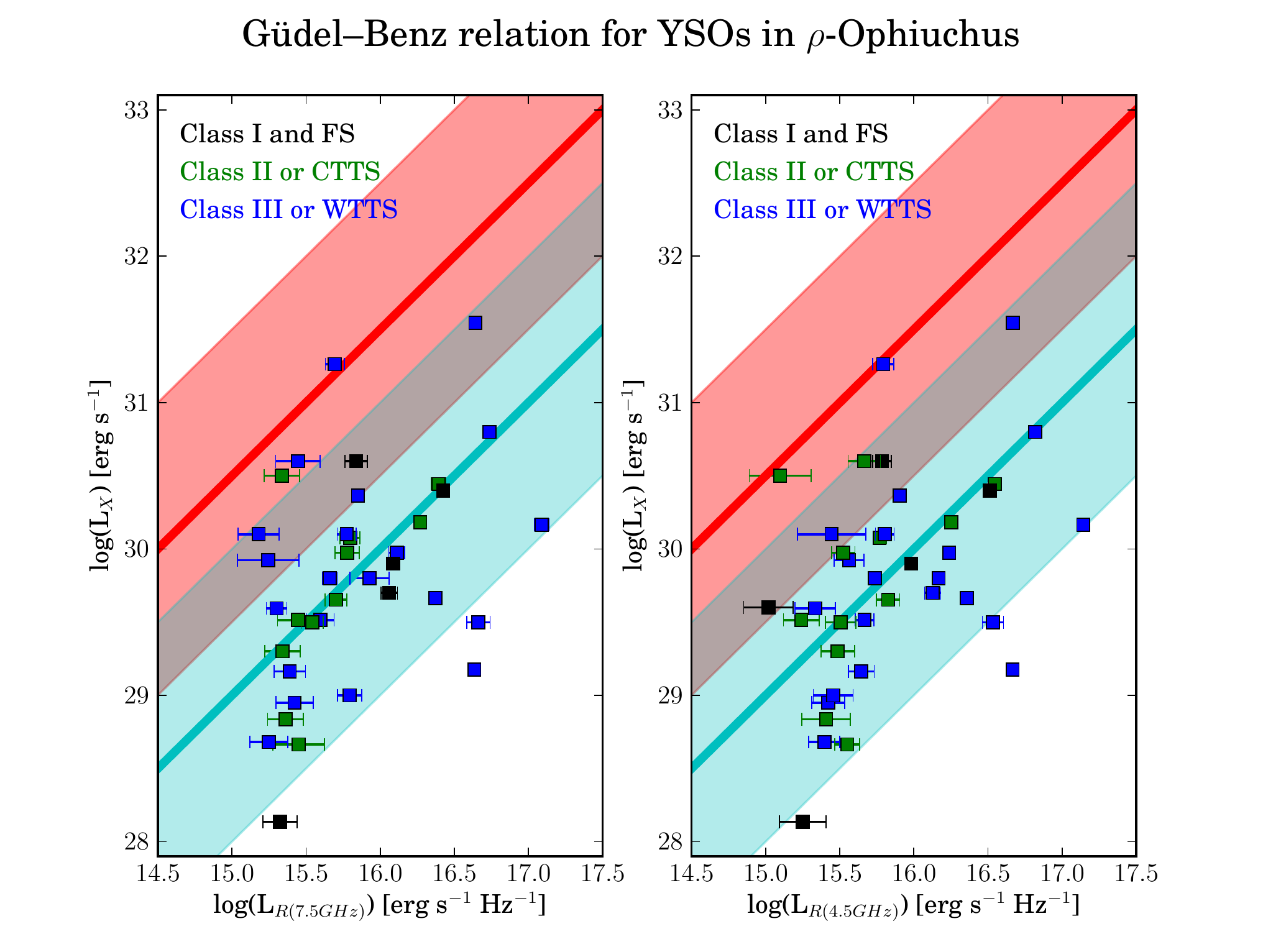}
\end{center}
\caption{X-ray luminosity as a function of radio luminosity. The red line corresponds to the 
G\"udel-Benz relation with $\kappa=1$ while the red strip centered on it represents a one
order of magnitude dispersion around that relation. The blue line and strip correspond to
the G\"udel-Benz relation but with $\kappa=0.03$. Colors indicate the evolutionary status 
of the object as explained at the top-left of the diagram.}
\label{fig:GB}
\end{figure*}

\begin{figure*}[b!]
\begin{center}
\includegraphics[width=0.65\textwidth,angle=0]{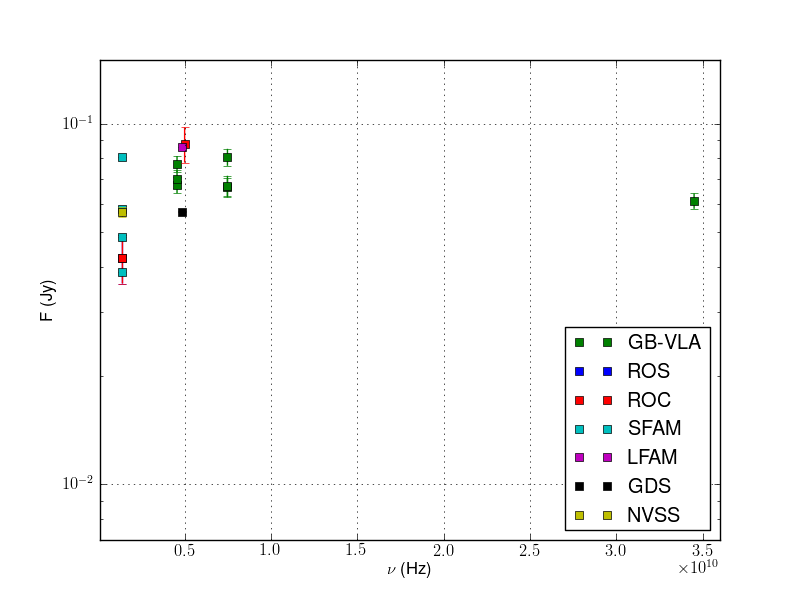}
\end{center}
\caption{Spectral energy distribution of the extragalactic source GBS-VLA J162700.00-242640.3.
The points correspond to the catalogs given in the legend (GBS-VLA correspond to the 
present observations, whereas the other symbols are the same as given in footnote $d$
of Table \ref{tab:counterparts}.}
\label{fig:calibrator}
\end{figure*}

\end{document}